\PassOptionsToPackage{pdfpagelabels=false}{hyperref}
\RequirePackage{rotating}
\documentclass[usenatbib]{mnras}
\usepackage{xcolor, graphicx}
\usepackage{amsmath}
\usepackage{amssymb}
\usepackage{soul}
\usepackage{txfonts}
\usepackage{journal_names}
\usepackage{caption}
\usepackage{todonotes}
\usepackage[export]{adjustbox}
\usepackage{longtable}
\usepackage{rotating}
\usepackage{lscape}

\newcommand{\Msun}{\rm M_\odot}
\newcommand{\kms}{\mbox{km s$^{-1}$}}
\newcommand{\vlos}{V_{\rm los}}
\newcommand{\SBunit}{\mbox{mag arcsec$^{-2}$}}
\newcommand{\re}{R_{\rm e}}

\newcommand{\unsim}{\mathord{\sim}}
\def\degr{\hbox{$^\circ$}}
\def\arcmin{\hbox{$^\prime$}}
\def\arcsec{\hbox{$^{\prime\prime}$}}

\def\Sersic/{{S\'ersic}}

\title[Low surface brightness galaxies in the Coma cluster]
{An expanded catalogue of low surface brightness galaxies in the Coma cluster using Subaru/Suprime-Cam}
\author[Alabi~et~al.~ ]
{Adebusola B. Alabi$^{1,3}$\thanks{Email: aalabi@ucsc.edu}, Aaron J. Romanowsky$^{1,2}$, Duncan A. Forbes$^{3}$,  Jean P. Brodie$^{1,3}$, 
\newauthor Nobuhiro Okabe$^{4,5,6}$\\
\\
$^{1}$ University of California Observatories, 1156 High Street, Santa Cruz, CA 95064, USA\\
$^{2}$ Department of Physics and Astronomy, San Jos\'e State University, San Jose, CA 95192, USA\\
$^{3}$ Centre for Astrophysics \& Supercomputing, Swinburne University, Hawthorn VIC 3122, Australia\\
$^{4}$ Department of Physical Science, Hiroshima University, 1-3-1, Kagamiyama, Higashi-Hiroshima, Hiroshima 739-8526, Japan\\
$^{5}$ Hiroshima Astrophysical Science Center, Hiroshima University, 1-3-1, Kagamiyama, Higashi-Hiroshima, Hiroshima 739-8526, Japan\\
$^{6}$ Core Research for Energetic Universe, Hiroshima University, 1-3-1, Kagamiyama, Higashi-Hiroshima, Hiroshima 739-8526, Japan\\
}

\begin{document}

\date{Accepted today}
\pagerange{\pageref{firstpage}--\pageref{lastpage}} \pubyear{2020}
\maketitle

\label{firstpage}
\begin{abstract}
We present a catalogue of low surface brightness (LSB) galaxies in the Coma cluster obtained from deep Subaru/Suprime-Cam $V$ and $R$-band imaging data within a region of $\unsim4$ deg$^{2}$. We increase the number of LSB galaxies presented in Yagi et al. (2016) by a factor of $\unsim3$ and report the discovery of $29$ new ultra-diffuse galaxies (UDGs). We compile the largest sample of ultra-diffuse galaxies with colours and structural parameters in the Coma cluster. While most UDGs lie along the red-sequence relation of the colour--magnitude diagram, $\unsim16$~per cent are outside (bluer or redder) the red-sequence region of Coma cluster galaxies. Our analyses show that there is no special distinction in the basic photometric parameters between UDGs and other LSB galaxies. We investigate the clustercentric colour distribution and find a remarkable transition at a projected radius of $\unsim0.6$~Mpc. Within this cluster core region and relative to the red-sequence of galaxies, LSB galaxies are on average redder than co-spatial higher surface brightness galaxies at the $2\sigma$ level, highlighting how vulnerable LSB galaxies are to the physical processes at play in the dense central region of the cluster. The position of the transition radius agrees with expectations from recent cosmological simulation of massive galaxy clusters within which ancient infalls are predicted to dominate the LSB galaxy population.
\end{abstract}


\section{Introduction}
There has been a rekindling of interest in low surface brightness (LSB) galaxies with the recent discovery of ultra-diffuse galaxies (UDGs) in the Coma cluster. UDGs are extended LSB galaxies with effective radii $\re \gtrsim 1.5$~kpc, central surface brightnesses $\mu(g,0) \gtrsim 24~\SBunit$, and exponential light profiles, i.e. \Sersic/ index, $n\unsim1$. The first $47$ UDGs were catalogued by \citet{vanDokkum_2015}, mostly outside the cluster core, using the remarkable Dragonfly Telephoto Array \citep{Abraham_2014}. Several groups have since reported the discovery of more UDGs in the Coma cluster \citep[e.g.,][]{Koda_2015, Yagi_2016,Ruiz_2018, Zaritsky_2019, Chilingarian_2019}. To date, 
$\unsim30$ Coma cluster UDGs have been spectroscopically confirmed as cluster members \citep{Kadowaki_2017, Alabi_2018, Ruiz_2018, Chilingarian_2019}.

While the debate about the true origin and nature of these enigmatic galaxies still persists in the literature, we note that most of the Coma cluster UDGs still lack colour information: a fundamental diagnostic in understanding the formation and evolution of galaxies \citep{Renzini_2006}. This omission may be directly linked to the fact that most of the UDG discoveries were made from analysis of single-band photometry and that faint photometry is challenging. Only $52$ UDGs out of the $854$ Coma cluster LSB galaxies in the catalogue of \citet{Yagi_2016} have $B-R$ colour measurements, and all within a $\unsim0.7$~deg$^2$ area\footnote{This is only $\unsim10$~per cent of the area defined by the projected virial radius of the Coma cluster, i.e. $\unsim2.9$~Mpc \citep{Kubo_2007}}. The faint LSB galaxy catalogue of \citet{Adami_2006} with $B-R$ colour covers a similarly small central $\unsim0.6$~deg$^2$ area within the Coma cluster. The recently published SMUDGes catalogue \citep{Zaritsky_2019}, which extends beyond the virial radius of the Coma cluster, has colour measurements available for only $43$ UDGs from the \citet{Yagi_2016} catalogue. The small sample size and limited radial coverage of Coma cluster UDGs with known colours therefore makes a systematic photometric study that is targeted at UDGs an urgent necessity.  

%why is this study necessary
While the environmental variation of galaxy colours with projected distance from the centre of the Coma cluster is well established in the literature for bright galaxies \citep{Terlevich_2001, Mahajan_2011} and dwarf galaxies \citep{Secker_1997}, the situation for UDGs, and in general, LSB galaxies within the Coma cluster, remains unclear. For example, \citet{Adami_2006} did not find any significant variation in colour with clustercentric radius in their faint LSB galaxy sample, perhaps due to the radially limited nature of their data. \citet{Terlevich_2001} attributed the blueing of mean galaxy colours with projected distance from the centre of the Coma cluster to an age effect, i.e., the cluster core is dominated by redder galaxies with older stellar ages, while the outskirts region is dominated by bluer galaxies with younger stellar ages. This is consistent with results from the spectroscopic studies of \citet{Smith_2009} and \citet{Smith_2011}, although \citet{Carter_2002} claimed that the environmental colour trends could also be a metallicity effect. \citet{Smith_2011} additionally showed that low mass galaxies have steeper radial age trends compared to their more massive counterparts. There are indications that UDGs (and by extension LSB galaxies) may have a similar clustercentric trends to dwarf galaxies \citep{Roman_2017, Alabi_2018, Ferre_2018, Mancera_2019, Zaritsky_2019, Chilingarian_2019} but the true behaviour within the Coma cluster is still unknown due to the paucity of UDGs with optical colour data.

Furthermore, while the vast majority of cluster UDGs are quiescent and lie on the red-sequence of the colour--magnitude relation \citep{Koda_2015}, a few UDGs with significant deviations from the red-sequence --mostly bluewards-- have been reported \citep{Roman_2017, Chilingarian_2019, Zaritsky_2019}, mostly in low density environments including the cluster outskirts \citep{Alabi_2018}. \citet{Adami_2006} previously identified faint LSB galaxies in the Coma cluster with colours redder or bluer than the red-sequence galaxies. They interpreted these faint LSB galaxies as recent cluster infalls from less--dense environments where they have been ``pre-processed'' to varying degrees \citep{Zabludoff_1998}. Identifying galaxies that are bluer or redder relative to the red-sequence region from coherent photometry is therefore useful in understanding the origin of present-day cluster LSB galaxies. 

%Solution from this work
In this work, we present a catalogue of galaxies within $\unsim4$~deg$^2$ of the Coma cluster. We probe the same region of the Coma cluster as in the $R$-band LSB galaxy studies of \citet{Koda_2015} and \citet{Yagi_2016}, hitherto the most comprehensive works on Coma cluster LSB galaxies, employing an additional $V$-band filter in order to obtain colours and securely discriminate against contamination from background galaxies. We present a description of our imaging data in Section~\ref{sec:data}. We give details of our data analysis in Sections~\ref{sec:obj_detect} and \ref{sec:gal_model} including galaxy detection, galaxy modelling with \texttt{GALFIT}, and removal of confirmed contaminants from our catalogue. We present our Coma cluster colour--magnitude diagram in Section~\ref{sec:cmd} and discuss the residual contamination in our final catalogue. Finally, we present our final catalogue and summarize our results in Section~\ref{sec:summ}. 

Throughout this work, we adopt a distance of $100$~Mpc, a redshift of $0.023$, a distance modulus of $(m-M)_{\rm 0} = 35$ \citep{Carter_2008}, a virial radius of $\unsim2.9$~Mpc \citep{Kubo_2007}, position angle of $71.5$~deg \citep{Plionis_1994}, and central co-ordinates RA: 12:59:48.75 and Dec: +27:58:50.9 (J2000) for the
Coma cluster. We also adopt the following cosmology: $\rm \Omega_m=0.3$, $\rm \Omega_{\Lambda}=0.7$ and, $\rm H_{0}=70\ \kms\ Mpc^{-1}$. 

\section{Data}
\label{sec:data}
\begin{figure}
    \includegraphics[width=0.48\textwidth]{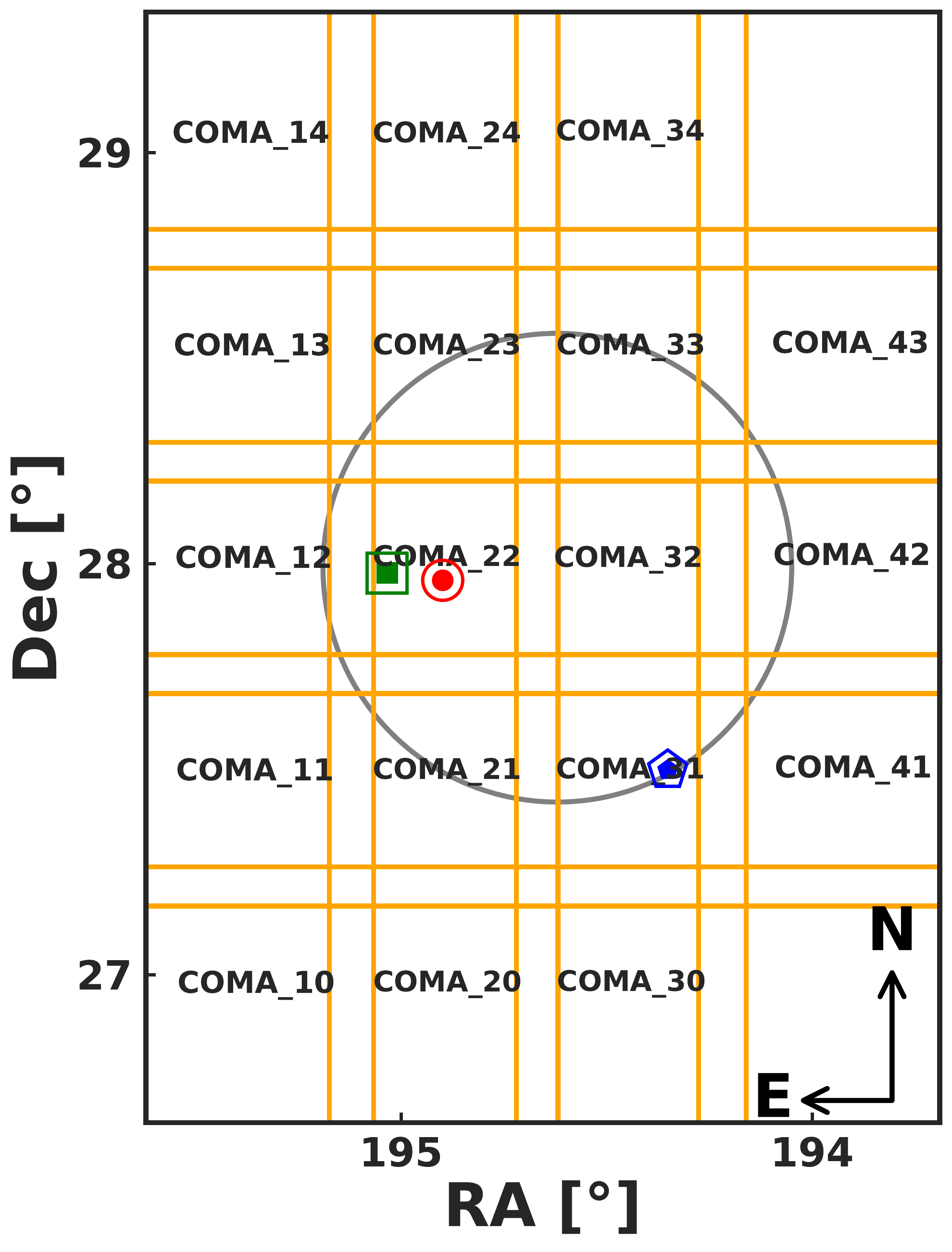}\hspace{0.01\textwidth}\\ 
	\caption{\label{fig:outline} Mosaic of the $18$ pointings used to observe the Coma cluster. The final mosaic covers $\unsim4.2\degr$ region of the Coma cluster. Galaxy IDs in the final catalogue are prefixed with the names of their originating pointing as shown in the plot. Multiple galaxy detections from the overlapping regions between adjacent pointings are used later in the text to quantify reliable uncertainties on the modelled galaxy structural parameters. As a guide to the physical scale, we show a circle with projected radius of $1$~Mpc within which we have marked the positions of the three cD galaxies: NGC~4889 (green square), NGC~4874 (red circle), and NGC~4839 (blue pentagon).}
\end{figure}

We analyse the Coma cluster Subaru/Suprime-Cam \citep{Miyazaki_2002} $V$ and $R$ band imaging data previously reduced and published in the weak-lensing study of \citet{Okabe_2014}. The imaging is made up of a mosaic of $18$ individual pointings with $2\arcmin$ overlap between adjacent fields and a total sky coverage of $\unsim4$~deg$^2$. Details of the exposure times and seeing per pointing can be found in \citet{Okabe_2014} alongside a summary of the data reduction process. Average exposure time and seeing FWHM are $14$~minutes $\& ~1\arcsec$ and $26$~minutes $\& ~0.7\arcsec$, in the $V$ and $R$-bands, respectively. We note that our $R$-band imaging is slightly different from that used by \hypertarget{Y16}{\citet[][hereafter \hyperlink{Y16}{Y16}]{Yagi_2016}}, where they included additional imaging data with worse seeing in $4$ pointings in their analysis. Figure~\ref{fig:outline} shows the mosaic of the $18$ pointings and the overlapping regions between them.

We used the publicly available code by \citet{Kelly_2014} to check and confirm the zero-point magnitudes originally used for photometric calibration by \citet{Okabe_2014} using bright stars in common with the Sloan Digital Sky Survey (SDSS) catalogue \citep{Ahn_2012}. Our zero-points are similar to those obtained by \citet{Okabe_2014}.
We apply Galactic extinction corrections on a per-pointing basis to the magnitude and surface brightness measurements, using values from the dust extinction maps of \citet{Schlafly_2011}. This is because extinction variation from object to object within the same pointing is very small compared to the estimated photometric uncertainties (see Section~\ref{sec:est_unc} for more details). The applied corrections vary from $0.021$--$0.033$~mag and $0.017$--$0.026$~mag in the $V$ and $R$ bands, respectively. We also applied ``K-corrections'' of $0.03$ and $0.02$ to $V$ and $R$-band photometry, respectively, determined using the K-corrections calculator \citep{Chilingarian_2010, Chilingarian_2012}. All magnitudes hereafter are in the AB system, and are extinction and K--corrected. 

%coma11,coma20,coma21,coma31

\section{Object Detection}
\label{sec:obj_detect}
We perform initial object detection on both $V$ and $R$-band images with \texttt{SExtractor} \citep{Bertin_1996}, adjusting the configuration criteria to maximize the detection of galaxies from the Coma cluster low surface brightness catalogue of \protect\hyperlink{Y16}{Y16}. We run \texttt{SExtractor} in the ``dual-image'' mode, using the $R$-band imaging for object detection and use the following criteria, similar to that used in \protect\hyperlink{Y16}{Y16}, to identify galaxy candidates in our \texttt{SExtractor} catalogue: 
\begin{itemize}
    \item non-zero \texttt{SExtractor} Petrosian RADIUS,
    \item \texttt{SExtractor} FWHM $\geq 5$~pixels ($\sim 0.5$~kpc at the distance of Coma cluster),
    \item \texttt{SExtractor} FLAG parameter $< 4$ (objects with defects such as saturated pixels or truncated isophotes are excluded),
    \item \texttt{SExtractor} CLASS\_STAR parameter $< 0.5$ (objects with CLASS\_STAR~$\sim1$ are foreground stars),
    \item $R$-band magnitude uncertainty $< 0.2$ (this corresponds to a minimum signal-to-noise ratio (SNR) $\sim5$).
\end{itemize}
This initial list of criteria help to exclude spurious detections, foreground stars, unresolved compact sources (globular clusters), and some background galaxies from our catalogue, reducing the size of our \texttt{SExtractor} source list by a factor of $\sim4$ to $\sim200,000$ objects. 

\begin{figure}
    \includegraphics[width=0.48\textwidth]{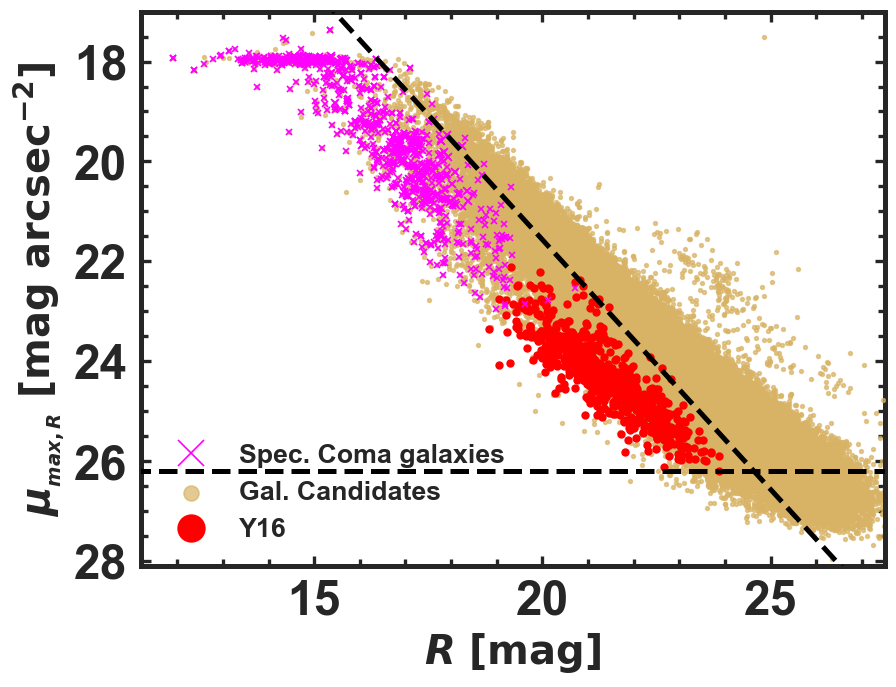}\hspace{0.01\textwidth}\\
	\caption{\label{fig:bkgrd_sep} Peak surface brightness--$R$-band magnitude diagram showing Coma cluster galaxy candidates (brown dots) from our \texttt{SExtractor} analysis. We also show spectroscopically confirmed Coma cluster galaxies from various sources in the literature (magenta X's). At any surface brightness, likely Coma cluster galaxies have brighter apparent magnitudes. The inclined line shows the expected peak surface brightness for a galaxy with $\re \sim 1.2\arcsec$ (the worst FWHM seeing in our imaging data equivalent to $\sim0.6$~kpc at the distance of Coma cluster), while the horizontal line is our faint peak surface brightness limit ($\unsim26.1$~\SBunit), defined using the low surface brightness catalogue of \citet[][Y16]{Yagi_2016} (red dots). Galaxy candidates with saturated central pixels around $\mu_{max,R}\sim18.1$~\SBunit are excluded from subsequent analysis. The magnitudes shown here and hereafter have all been corrected for Galactic extinction and are K-corrected.}
\end{figure}

Next, we use the peak surface brightness -- magnitude diagram as a diagnostic to further remove contaminants from our sample of galaxy candidates. As shown in Figure~\ref{fig:bkgrd_sep}, galaxies that belong to the Coma cluster tend to occupy a distinct region compared to the fainter background galaxies. We compile and show a sample of confirmed Coma cluster galaxies from the literature which we match to our object catalogue. These galaxies vary from giants to dwarf galaxies and are selected from the spectroscopic studies of \citet{Mobasher_2001}, \citet{Edwards_2002}, \citet{Aguerri_2005}, \citet{Smith_2009}, the NASA/IPAC Extragalactic Database (NED)\footnote{http://ned.ipac.caltech.edu/} Coma cluster galaxy list, and SDSS. The compilation, however, contains only bright galaxies, i.e. $R\leq19$~mag, which have been the traditional subjects of spectroscopic studies. We extend this sample to fainter magnitudes ($R\sim24$~mag) by including the LSB galaxies from \hyperlink{Y16}{Y16}, which have been shown to be mostly cluster members in various studies \citep[e.g.][]{Kadowaki_2017, Alabi_2018, RL_2018, Chilingarian_2019}.

The inclined line shows the expected peak surface brightness for a galaxy with $\re \sim 1.2\arcsec$ assuming an exponential profile, i.e. \Sersic/ index, $n=1$, (the worst FWHM seeing in the $V$-band equivalent to $0.6$~kpc at the distance of the Coma Cluster) at various magnitudes. We therefore exclude galaxy detections smaller than this size-limit from subsequent analysis, since they are most likely background sources, ultracompact dwarfs, or imaging defects such as false detections in the halos of bright stars. Likewise, we use the LSB catalogue of \protect\hyperlink{Y16}{Y16} to define the faint surface brightness limit of our catalogue. We adopt a peak surface brightness limit of $\mu_{max,R}\sim26.1$~\SBunit~which is equivalent to a mean effective surface brightness of $27.2$~\SBunit. We also exclude all galaxy candidates ($209$) with saturated central pixels seen around $\mu_{max,R}\sim18.1$~\SBunit. With the application of these thresholds, the size of our catalogue is now reduced to $27,437$ galaxies which is better suited for subsequent \texttt{GALFIT} analysis, including $1,305$ galaxies detected in more than one pointing. Likewise, out of the $854$ LSB galaxies published in \protect\hyperlink{Y16}{Y16}, our catalogue contains $757$ galaxies with detection in both bands, alongside $181$ duplicate detections. We later use these duplicate detections to estimate the \textit{true} uncertainties on the structural parameters of our galaxies.

\begin{figure}
    \includegraphics[width=0.48\textwidth]{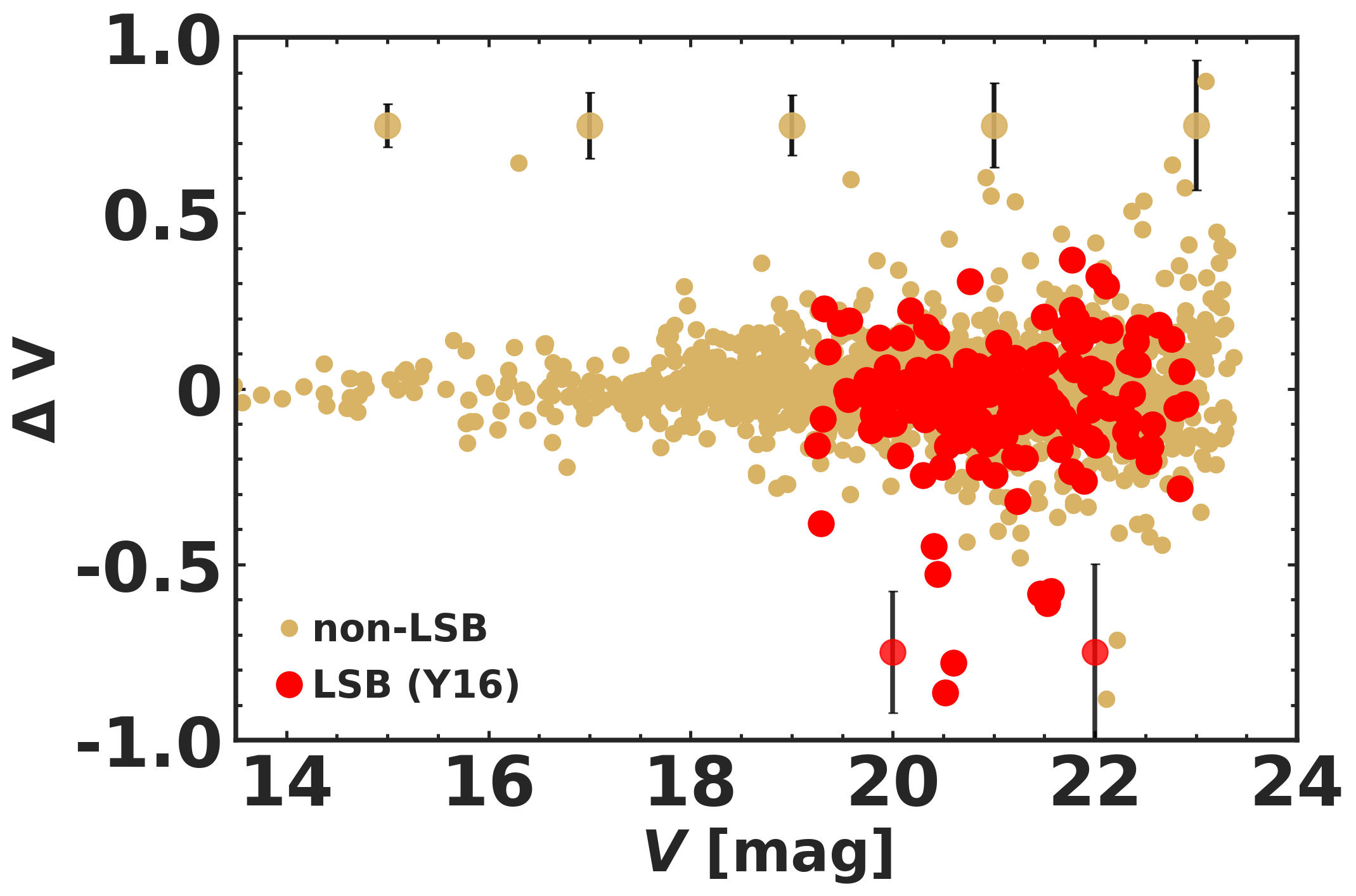}\hspace{0.01\textwidth}\\ 
	\caption{\label{fig:duplicates} Variation in \texttt{GALFIT} $V$-band magnitude measurements in the $1,305$ galaxies with duplicate detections. We show the low surface brightness galaxies from the catalogue of \protect\hyperlink{Y16}{Y16} as red circles, while the remaining cluster galaxies are shown as smaller brown circles. The corresponding error-bars are the standard deviations determined in magnitude bins of width $\Delta V = 2$~mag  and they increase up to a maximum of $\sim0.5$ at very faint magnitudes.}
\end{figure}

\section{Galaxy modelling with GALFIT}
\label{sec:gal_model}
The projected intensity profile, $I(R)$, of a galaxy can be described with a \Sersic/ function \citep{Sersic_1968,Graham_2005} of the form: 
\begin{equation}
I(R) = I_{e}\exp\left[-b_{n}\left(\left(\frac{R}{R_e}\right)^{1/n} - 1\right)\right]
\end{equation}
where $I_{e}$ is the intensity at the effective radius $\re$ which encloses half the total galaxy luminosity, $n$ is the \Sersic/ index, also known as the profile shape parameter, and $b_{n}$ is a complicated function of $n$. Bright elliptical galaxies which are centrally concentrated typically follow de Vaucouleurs profiles \citep{deV_1948, deV_1959} with $n\sim4$, while the less centrally concentrated spiral disks and LSB galaxies have exponential profiles with $n\sim1$. 

To obtain the structural properties of all the galaxies in our catalogue, we fit \Sersic/ functions in both $V$ and $R$ bands independently with \texttt{GALFIT}\footnote{http://users.obs.carnegiescience.edu/peng/work/galfit} \citep{Peng_2010}. We make postage-stamp cutouts for each galaxy candidate using the positions and size estimates from our initial \texttt{SExtractor} analysis as a basis. Each cutout is centred on the central pixel coordinates from \texttt{SExtractor} with dimensions set to $10$ times the $\re$ to allow for good sky estimation. We then identify and mask out all bright sources within the frame that do not belong to the target galaxy in an automated way. We jointly fit for the position, magnitude, $\re$, $n$, axis ratio ($q$), and position angle ($PA$) of the target galaxy, as well as the sky background value. Point-spread functions (PSF) used in the fitting process are constructed for each pointing from a sample of bright, unsaturated stars (on average, we use $\sim50$ stars per pointing), pre-selected from Section~\ref{sec:obj_detect}. %Figure~\ref{fig:gal_fit} shows various galaxy images and their best-fit \texttt{GALFIT} models over a wide range of magnitudes and mean effective surface brightness.

\subsection{Quality Control}
\label{sec:qty_ctrl}
In order to identify galaxies with reliable structural parameters, we compare the formal uncertainties on the magnitudes obtained from \texttt{GALFIT} with estimates of their systematic errors. We obtain these systematic errors from the sample of galaxies with duplicate observations and require that for a good fit, the formal uncertainties should be less than the systematic errors \citep[][]{Haussler_2007}. This requirement implies that the uncertainties on the magnitude measurements are never dominated by pixel noise.

Figure~\ref{fig:duplicates} shows how $V$-band magnitude measurements differ in galaxies with duplicate detections. At fainter magnitudes and in the LSB galaxies, where SNR is significantly reduced and galaxy edges are difficult to identify, the differences between duplicate measurements are significantly increased, reaching $\unsim0.5$~mag. We use the standard deviations of these distributions, i.e., $\sigma_{\Delta V}\unsim0.16$; $\sigma_{\Delta R}\unsim0.20$, as limits in determining good and acceptable fits. This, in addition to a visual examination of the fits, reduces the size of our Coma cluster catalogue to $11,496$ galaxies. A visual examination of all the discarded galaxy candidates reveal that most of them are either spurious detections, typically, with extremely large $n$ values and smaller $\re$ or faint sources in heavily crowded fields.

\subsection{Correction to total magnitudes}
\label{sec:mag_corr}
Since the galaxy--background boundary is not as well-defined in LSB galaxies compared to their HSB counterparts \citep[e.g.][]{Trujillo_2001, Adami_2006, Haussler_2007, Wittmann_2017}, we investigate how robust the total magnitudes from \texttt{GALFIT} analysis are by also obtaining their curve-of-growth (COG) total magnitudes. We fix the position, $q$, and $PA$ to parameter values from our previous \texttt{GALFIT} analysis and estimate the total magnitudes within consecutive isophotes, starting from $0.25\re$, and moving outwards in steps of $3$~pixels, i.e., $0.3$~kpc, stopping when the magnitude difference between successive isophotes is $<0.02$~mag. We exclude galaxies that have nearby neighbors within a radius of $7\re$ from this analysis. 

\begin{figure}
    \includegraphics[width=0.48\textwidth]{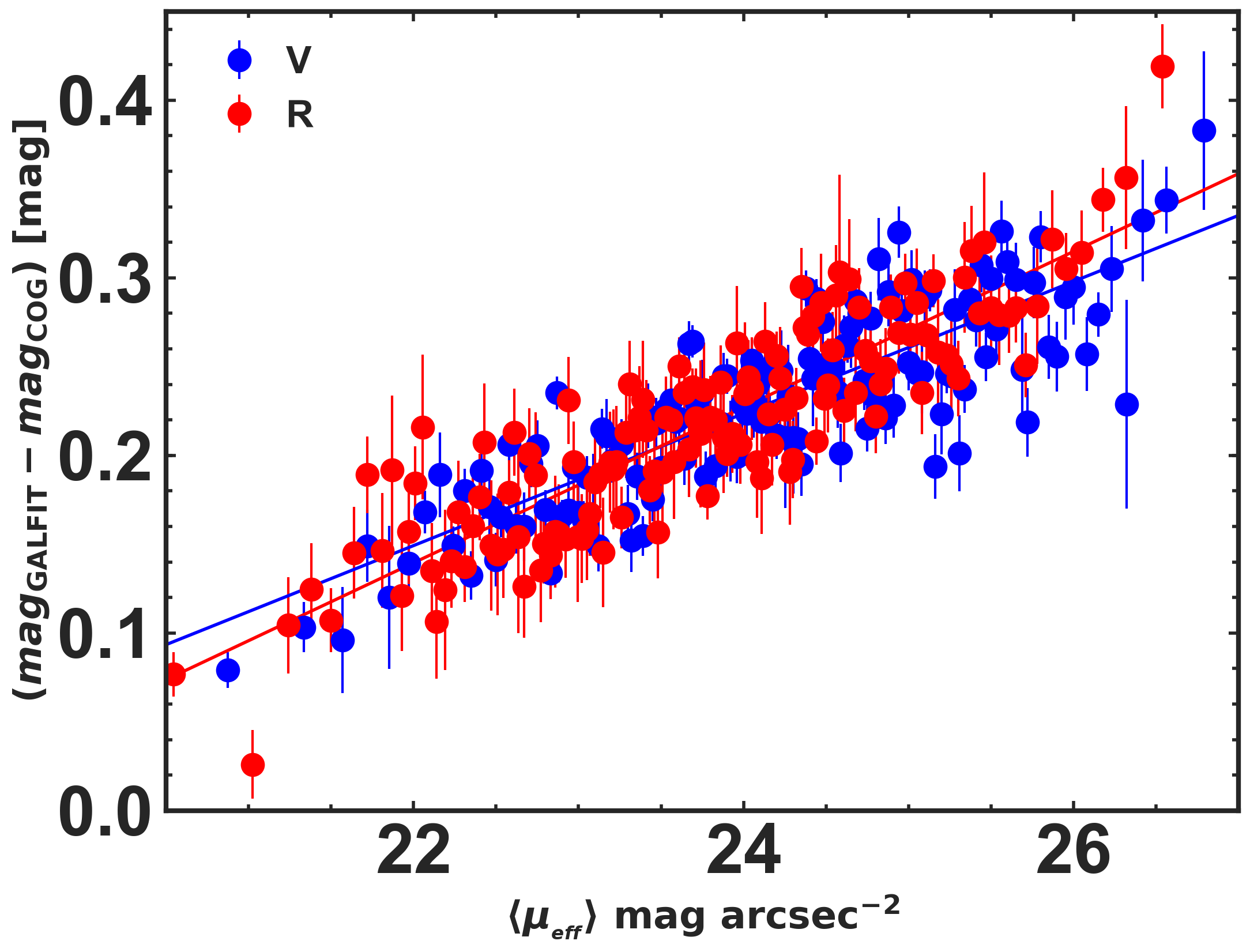}\hspace{0.01\textwidth}\\ 
	\caption{\label{fig:mag_corr} Mean offset between total magnitudes from \texttt{GALFIT} and asymptotic magnitudes from curve-of-growth analysis as a function of mean effective surface brightness for galaxies in our sample that are relatively isolated. The magnitude correction factor increases as galaxies become fainter, reaching a maximum of $\unsim0.4$~mag at $\unsim27$~\SBunit, where the galaxy--background boundaries become vague.}
\end{figure}

This exercise shows that the total integrated magnitude from \texttt{GALFIT} is systematically fainter than the asymptotic magnitude from the COG analysis, with the corrections becoming significant in both $V$ and $R$ bands, i.e., $\geq 0.2$~mag, only when the mean galaxy surface brightness is fainter than $\unsim24$~\SBunit. We show the mean magnitude corrections as a function of mean effective surface brightness in Figure~\ref{fig:mag_corr} and fit linear functions which we apply to total magnitudes from \texttt{GALFIT} analysis.  

\begin{align}    
\begin{aligned}
    \Delta V &= 0.037 \times \langle \mu_{\rm eff,V} \rangle - 0.67 \\
    \Delta R &= 0.044 \times \langle \mu_{\rm eff,R} \rangle - 0.82
\end{aligned}
\label{eq:mag_corr}
\end{align}

\subsection{Estimation of true uncertainties on structural parameters}
\label{sec:est_unc}
As already mentioned in Section~\ref{sec:qty_ctrl}, the formal errors from \texttt{GALFIT} are significantly lower than the systematic errors obtained from duplicate detections, at least for the total magnitudes. The formal uncertainties are based entirely on Poisson pixel noise and may be unrealistic estimates in cases such as when the adopted model does not adequately describe the galaxy light profile or when the boundaries of galaxies are difficult to identify due to low SNR, rapidly varying sky background, or crowding from nearby neighbours. 

\begin{figure*}
    \includegraphics[width=0.96\textwidth]{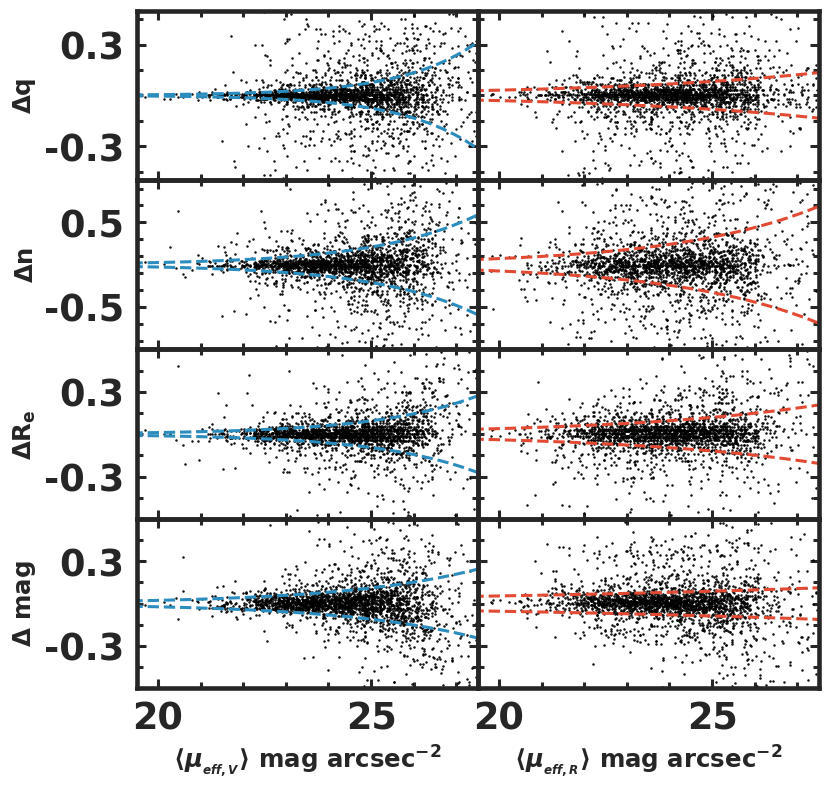}\hspace{0.01\textwidth}\\
	\caption{\label{fig:err_est} Variation of the structural parameters obtained from \texttt{GALFIT} modelling of galaxies with duplicate detections. We show on the x-axis $V$ and $R$-band mean surface brightness within the effective radius (\textit{left-hand panels} and \textit{right-hand panels}, respectively) and on the y-axis, magnitude, effective radius, \Sersic/ index, and axial ratio. The dashed lines are fits to the 68th percentile of the corresponding ordinate parameter (in log) and are used to estimate the true model uncertainties as described in the text.}
\end{figure*}

Figure~\ref{fig:err_est} shows how the deviations between repeat measurements of model parameters for the same galaxies vary as a function of mean surface brightness in both $V$ and $R$-bands. We fit linear functions of the form 
\begin{equation}
    {\rm log}~\sigma = \alpha \times \langle \mu_{\rm eff} \rangle + \beta
\label{eq:err_est}
\end{equation}
to the distribution of the 68th percentile of the difference of each model parameter in bins of mean surface brightness, with $\alpha$ and $\beta$ being the fit coefficients, respectively. The best-fitting $\alpha$ and $\beta$ coefficients are summarized in Table~\ref{tab:coeff} and are used to compute $\sigma$ for any measured mean surface brightness. Our final adopted error estimates, shown in Table~\ref{tab:complete_tab}, are obtained by adding $\sigma$ in quadrature to the \texttt{GALFIT} formal uncertainties.

\begin{table}
\centering
\begin{tabular}{@{}l c c c c}
\hline
\hline
  &  \multicolumn{2}{c}{$V$} & \multicolumn{2}{c}{$R$}\\
\hline
Parameter & $\alpha$ &  $\beta$  &  $\alpha$   &  $\beta$ \\
\hline
  mag 		& $0.137$ & $-4.393$ & $0.042$ & $-2.11$ \\
  $\re$ 	& $0.183$ & $-5.603$ & $0.096$ & $-3.326$ \\
  \Sersic/~$n$  & $0.18$ & $-5.178$  & $0.129$ & $-3.734$ \\
  $q$ 		& $0.248$  & $-4.318$ & $0.085$ & $-3.219$ \\
  $PA$ 	 	& $0.299$ & $-4.438$ & $0.189$ & $-3.831$ \\
\hline
\end{tabular}
\caption{Summary of the coefficients used in eq. \ref{eq:err_est} to estimate the systematic errors unaccounted for in the \texttt{GALFIT} analysis.}
\label{tab:coeff} 
\end{table}

\subsection{Removal of spectroscopic contaminants from the Cluster catalogue}
\label{sec:fin_clean}
\begin{figure*}
    \includegraphics[width=0.94\textwidth]{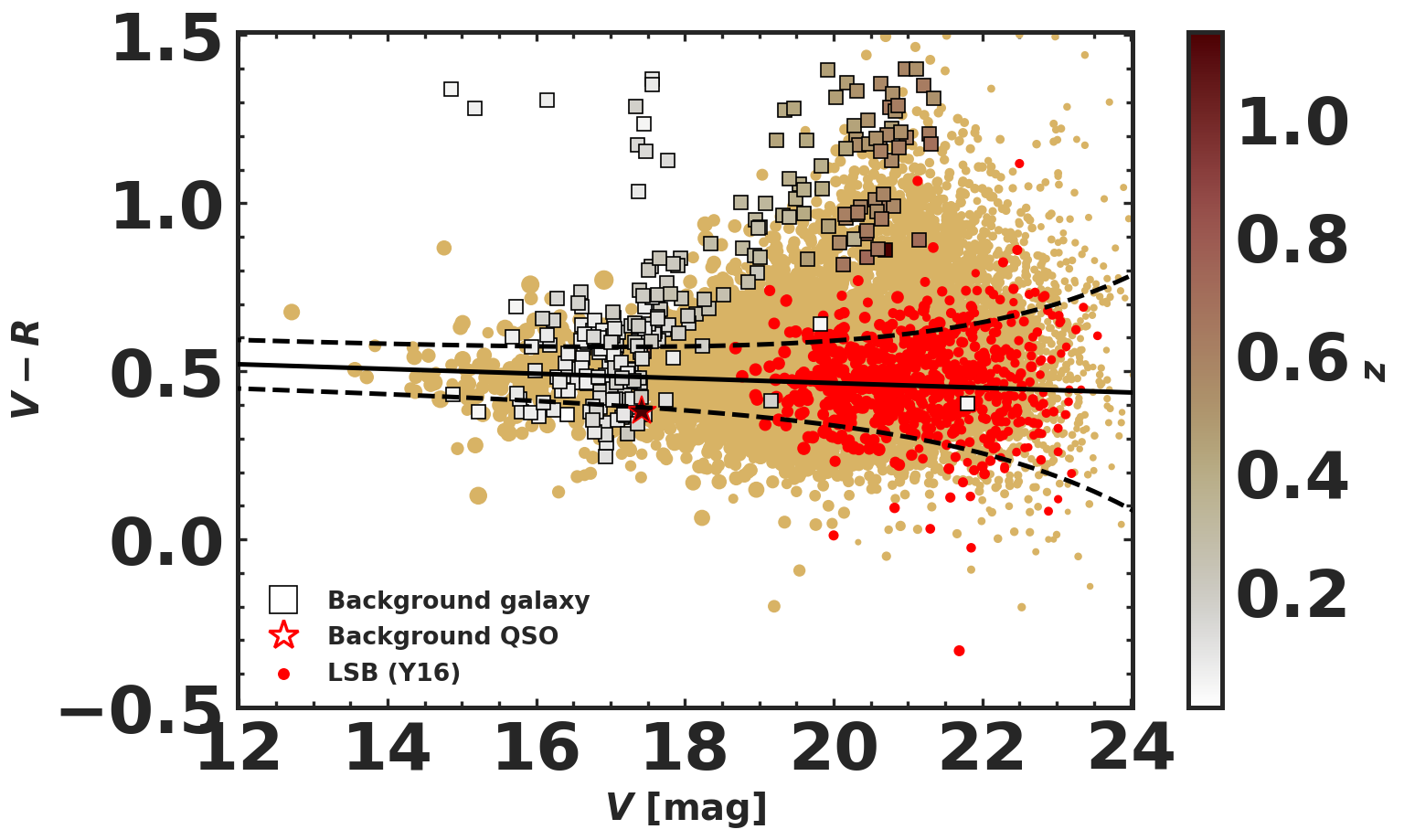}\hspace{0.01\textwidth}\\
	\caption{\label{fig:CMD} Colour--magnitude diagram of Coma cluster galaxies (brown circles) detected and analysed in this work. Galaxies from the low surface brightness catalogue of \protect\hyperlink{Y16}{Y16}, which mostly belong to the Coma cluster, are shown as red circles. Spectroscopically confirmed background galaxies in the direction of Coma cluster from the literature (filled squares) have been colour-coded by their redshifts. Also shown is a spectroscopically confirmed background quasi-stellar object (star marker with red edges, colour-coded by its redshift). We overlay the red-sequence (black solid line) obtained by a linear fit to the bright, spectroscopically confirmed cluster galaxies, extrapolated to faint magnitudes. We also show the $1\sigma$ intrinsic scatter  (black, dashed lines) around the red-sequence. The intrinsic scatter increases from $\unsim0.06$~mag at $V=15$~mag to $\unsim0.28$~mag for the faintest LSB galaxies. Most of the low surface brightness galaxies from \protect\hyperlink{Y16}{Y16} have colours consistent with the red-sequence and fall within the limits of the $1\sigma$ intrinsic scatter.}
	
    \includegraphics[width=0.9\textwidth]{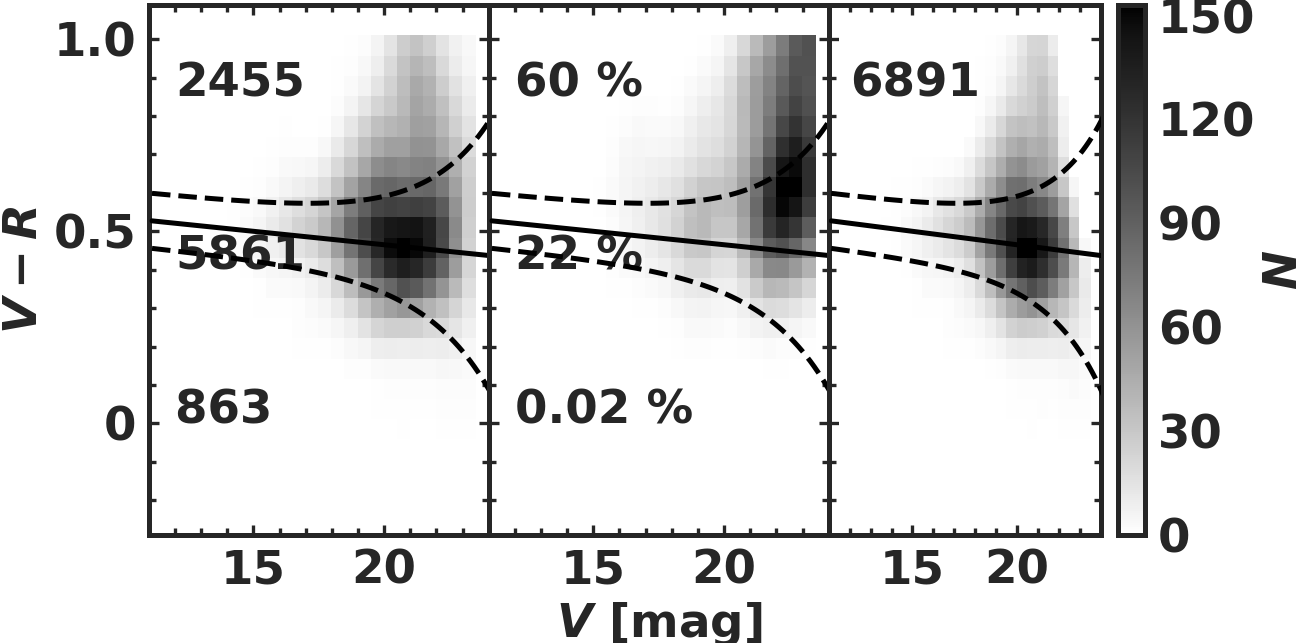}\hspace{0.01\textwidth}
	\caption{\label{fig:CMD_bkgrd_II} Histograms of colour--magnitude distribution of galaxies in the direction of the Coma cluster (\textit{Left panel}), and in a control field -- the Subaru Deep Field (\textit{Middle Panel}), overlaid with the red-sequence of known Coma cluster galaxies. The solid and dashed black lines in all the panels are the same as in Figure~\ref{fig:CMD}. The background contaminants are mostly faint galaxies with $V\unsim22.5$~mag and have very red colours $V-R\unsim0.7$. We summarize the number of galaxy candidates on and off the red-sequence in the left panel and the corresponding contamination levels in the middle panel. The \textit{Right Panel} shows the colour--magnitude distribution of the remaining $6,891$ Coma cluster galaxies after statistically correcting for the background contaminants. The colour-bar is the number density of the galaxies.}
\end{figure*}

As a final step in minimizing contaminants in our final catalogue, we retrieve spectroscopically confirmed foreground stars and background galaxies (including quasi-stellar objects (QSO)) in the direction of the Coma cluster from the literature. This compilation comes mostly from SDSS and NED ($266$ foreground stars, $208$ QSOs, and $1042$ background galaxies) supplemented with $31$ and $20$ background galaxies from \citet{Adami_2009} and \citet{Chiboucas_2010}, respectively. We also include $2$ galaxies from the \hyperlink{Y16}{Y16} LSB catalogue known to be background galaxies \citep{Kadowaki_2017, Alabi_2018}. This sample, which we compare with our catalogue, is such that the foreground stars have $-300 \leq \vlos\ [\kms] \leq300$, the QSOs $\vlos \geq 30000$~$\kms$, while the background galaxies have $\vlos \geq 11000$~$\kms$, all with photometric $V$-band magnitudes ranging from $14.5$ to $22$. While our catalogue is devoid of all the foreground spectroscopic stars, we find a QSO\footnote{This QSO is SDSS\ J125712.28+280543.3 at $z=1.29$ with $V$-band $\re\unsim5.4\arcsec$.} with $17.4\ V$--band magnitude as well as $251$ background galaxies in our catalogue as shown in Figure~\ref{fig:CMD}. Half of these spectroscopic background galaxies were catalogued as Coma cluster galaxy candidates in \citet{GMP_1983}. Appendix~\ref{sec:lit_cmp} contains an extensive comparison of our structural parameters with the literature after removing all the known contaminants.

\section{colour--magnitude diagram}
\label{sec:cmd}
We show the colour--magnitude diagram (CMD) of all galaxies with detections in both $V$ and $R$-bands in Figure~\ref{fig:CMD}. The $V-R$ colours shown here (and used in all subsequent analyses) are from the \texttt{SExtractor} dual-image mode analysis described in Section~\ref{sec:obj_detect}. We measure the colours within matched apertures based on $2.5~\times$ the Kron radius \citep{Kron_1980} parameter determined from the $R$-band imaging rather than from the $V$ and $R$ total magnitudes. We note that these matched aperture colours are generally $0.2-0.3$~mag redder than colours obtained from the total magnitudes. The black solid line, obtained from a linear fit to the bright ($13 \leq V\ \rm{[mag]} \leq 18$) and spectroscopically confirmed Coma cluster galaxies in our catalogue, highlights the red-sequence of galaxies. We extrapolate the red-sequence to fainter magnitudes and show in Figure~\ref{fig:CMD} that the faint LSB galaxies from the \protect\hyperlink{Y16}{Y16} are mostly consistent with the red-sequence, in agreement with a previous result from \citet{Koda_2015}. The $1\sigma$ intrinsic scatter around the red-sequence was obtained in magnitude bins with width $2$~mag and is shown as dashed lines in Figure~\ref{fig:CMD}. The scatter increases from $\unsim0.06$~mag at $V=15$~mag to $\unsim0.28$~mag for the faintest LSB galaxies. For subsequent analyses and discussion, we define three regions in our CMD: galaxies with colours within the limits defined by the $1\sigma$ scatter around the best-fit line to red-sequence galaxies (RSG); galaxies with colours redder than the $1\sigma$ limit ($>$RSG), i.e. redder than the average RSG; and galaxies with colors bluer than the $1\sigma$ limit ($<$RSG), i.e. bluer than the average RSG.

In the context of cluster photometry, the red-sequence is normally expected to be populated by quiescent galaxies that are bound to the host cluster. However, our analysis in Section~\ref{sec:fin_clean} shows that $\unsim10$~per cent of the RSG with magnitudes brighter than $V \unsim 19$~mag do not belong to the Coma cluster. These bright background galaxies which have redshifts $z\unsim0.2$ are known in the literature to be notoriously difficult to isolate with photometry only \citep{Adami_2006, Adami_2009, Mahajan_2011}. As earlier mentioned, we have excluded these background galaxies from our catalogue and are left with $1,564$ cluster galaxies with $V < 19$~mag. At fainter apparent magnitudes, i.e., $19 \leq V \rm{\ [mag]} \leq 22$, spectroscopic background galaxies in the direction of the Coma cluster from the literature have elevated redshifts $z \geq 0.3$ and are almost exclusively $>$RSG. A careful examination of these fainter background galaxies reveal that they generally have \Sersic/ index, $n>2$. We assume that all other faint $>$RSG galaxies with $n>2$ are background galaxies and therefore exclude them from our catalogue. This brings the number of remaining galaxies in our catalogue to $9,179$, with $64$, $27$, and $9$ per cent being RSG, $>$RSG, and $<$RSG, respectively.

\subsection{Control field}
\label{sec:SDF}
It should be noted that the spectroscopic sample used to isolate background galaxy contaminants is limited to high surface brightness galaxies, and as we have already shown, the red-sequence and surface brightness--magnitude diagnostics are both inadequate for weeding out all background galaxies. We therefore seek to know, statistically, the residual contamination level in our final catalogue. To this end, we analyse the $V$ and $R$ band observation of the Subaru Deep Field \citep[SDF;][]{Kashikawa_2004} as a control field. The SDF is at a similar Galactic latitude as the Coma cluster and was observed under similar photometric conditions (FWHM seeing $\unsim0.98\arcsec$). It has similar field-of-view and limiting surface brightness as each of our Coma pointings, although it reaches a limiting magnitude that is $\unsim1.5$~mag deeper than our Coma imaging data. Applying the same galaxy detection and quality control constraints (Sections~\ref{sec:obj_detect} \& ~\ref{sec:cmd}) to the control field and assuming that it is representative of our Coma fields, we expect $2,840$ background galaxies within our combined Coma fields, mostly at fainter magnitudes, i.e., $V \unsim 22$~mag, and redder colours, i.e., $V-R \unsim 0.7$, as shown in Figure~\ref{fig:CMD_bkgrd_II}.

Our control field analysis suggests that a minimum contamination level of $\unsim0.02$ per cent is realised in our photometric catalogue below the red-sequence ($<$RSG). Along the red-sequence, i.e. in the RSG group, the residual contamination is $\unsim20$ per cent, mostly among the fainter ($V > 19$~mag)  LSB galaxies. Above the red-sequence ($>$RSG), the residual contamination level is very high, reaching $\unsim60$ per cent. We summarize the residual contamination levels along and off the red-sequence in Figure~\ref{fig:CMD_bkgrd_II} and note that the total number of Coma cluster galaxies in our catalogue drops to $6,891$ galaxies after applying the statistical background-galaxy correction. 
 
\begin{figure*}
    \includegraphics[width=0.64\textwidth]{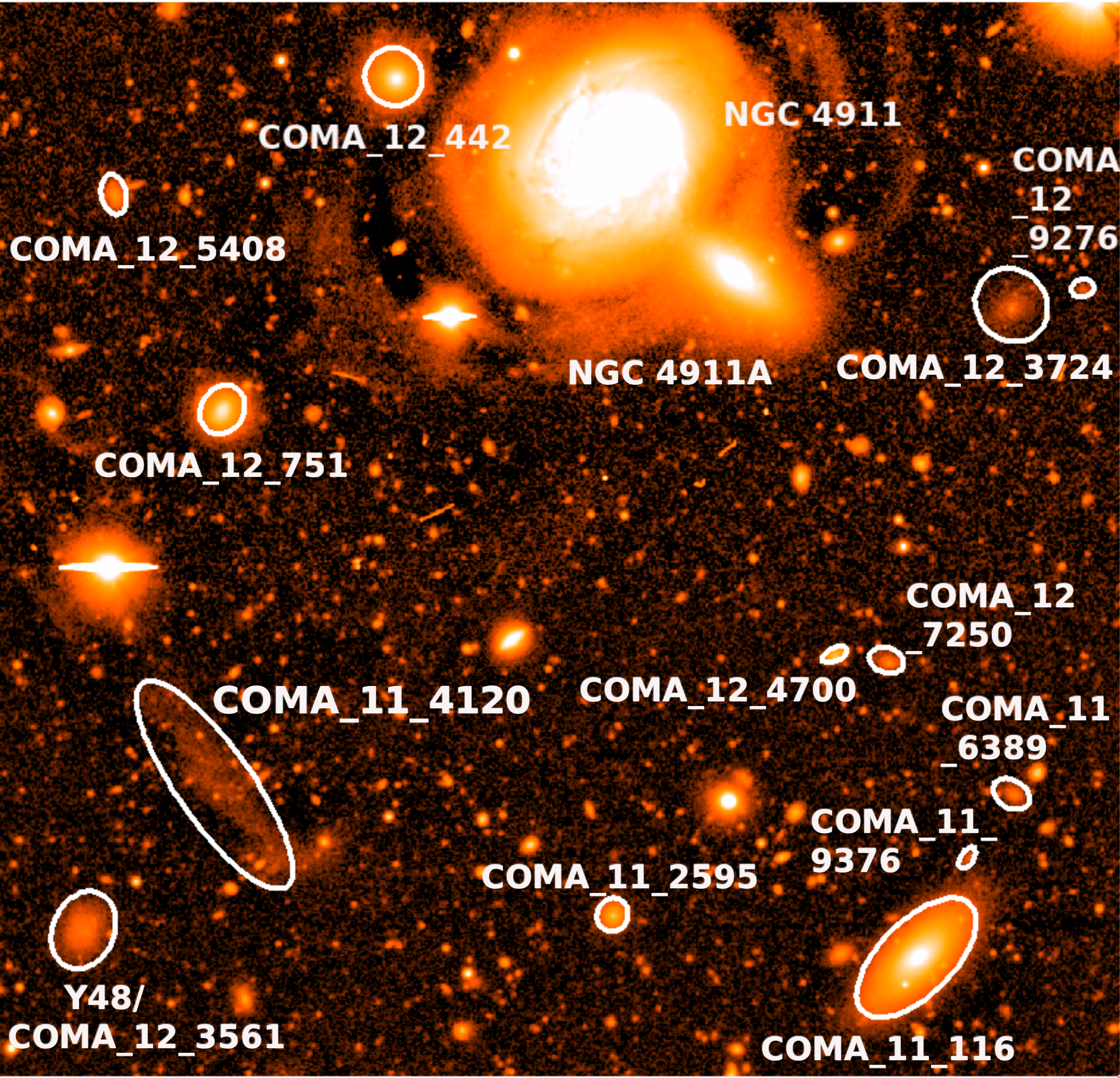}\hspace{0.01\textwidth}\\
	\caption{\label{fig:NGC4911} An example of the wide range of galaxies with structural parameters present in our final catalogue. We show the $V$-band Subaru/Suprime-Cam imaging of a field at $\unsim0.6$~Mpc from the centre of the Coma cluster with dimensions $120$~kpc~$\times$~$120$~kpc. The galaxies in our final catalogue are marked with white $1\re$ isophotes. This field contains the giant spiral galaxy, NGC~4911, several dwarf galaxies, and the ultra-diffuse galaxy, Y48 ($\rm COMA\_12\_3561$). It also includes a faint, disrupting galaxy ($\rm COMA\_11\_4120$) that is $\unsim75$~kpc away from the NGC~4911--NGC~4911A interacting pair. North is up and East is left.}
    \includegraphics[width=0.44\textwidth]{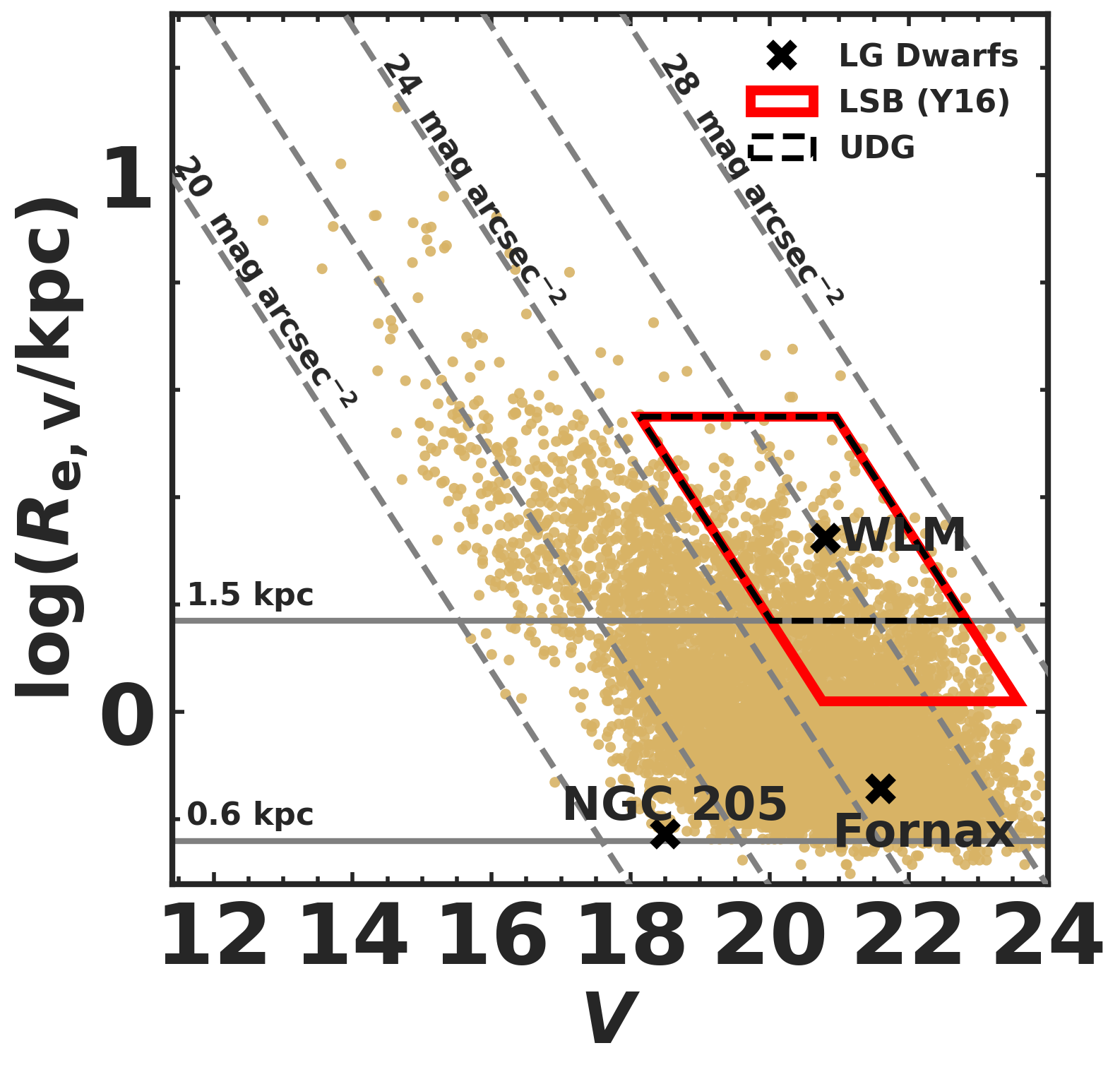}\hspace{0.01\textwidth}
    \includegraphics[width=0.46\textwidth]{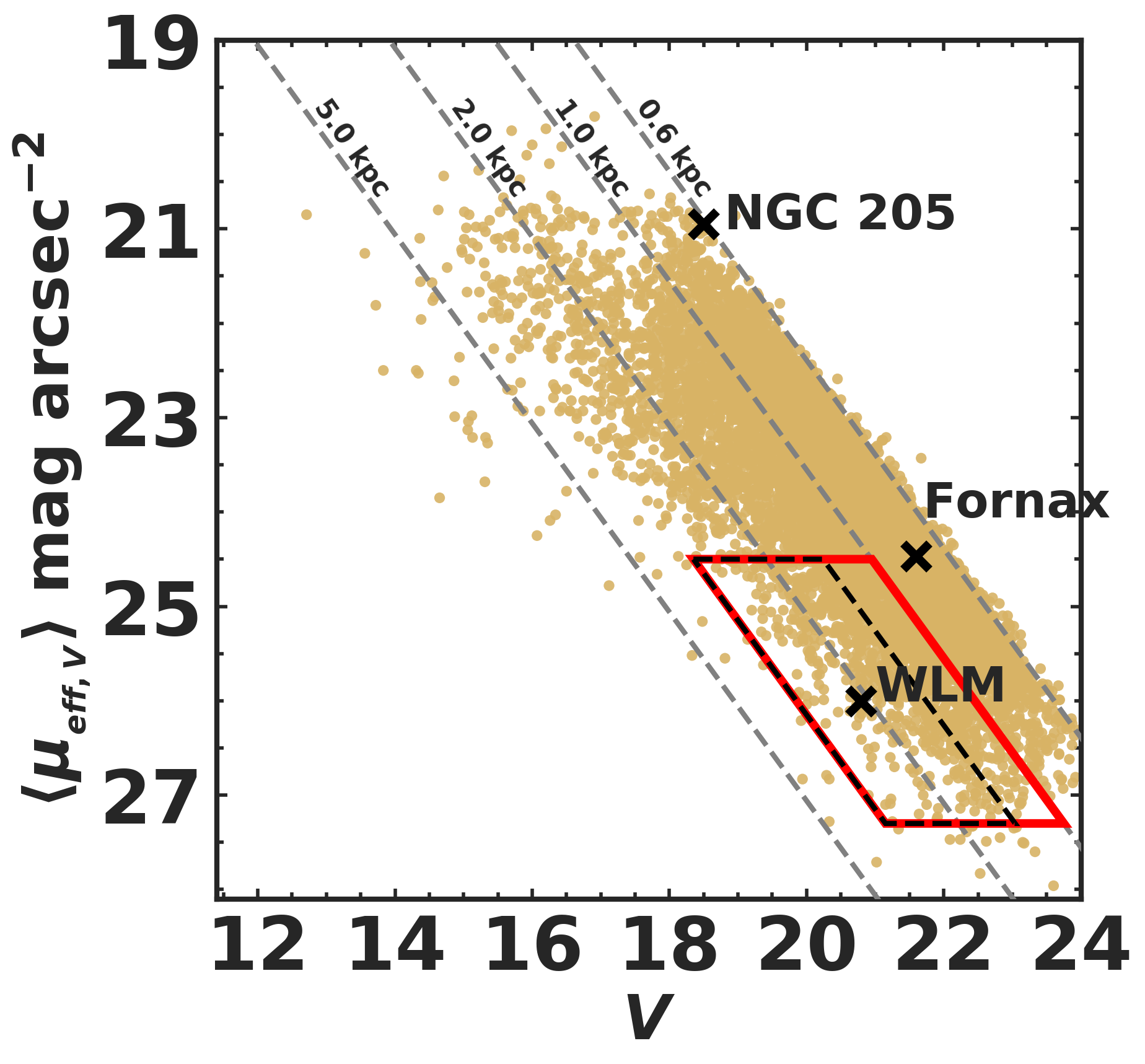}\hspace{0.01\textwidth}\\
	\caption{\label{fig:summ_cat} \textit{Left panel}: Size--magnitude distribution of Coma cluster galaxies. The brown circles are the $9,179$ galaxies successfully modeled with a single \Sersic/ function as described in the text. The dashed lines correspond to constant mean effective surface brightness. We detect Coma cluster galaxies as small as $\unsim0.6$~kpc and with mean surface brightness within the effective radius as faint as $\unsim27.5$~\SBunit. For context, we show the regions where the low surface brightness galaxies in the catalogue of \citet{Yagi_2016} may be found (red parallelogram), as well as that of the ultra-diffuse galaxies (black, dashed parallelogram). We also show and label some Local Group dwarf galaxies (black crosses) from the compilation of \citet{McConnachie_2012} that would fall within our detection limits, assuming they were observed at the distance of the Coma cluster. These are the dwarf irregular galaxy Wolf-Lundmark-Melotte (WLM), the Fornax dwarf spheroidal galaxy and the dwarf elliptical galaxy NGC~205.
	 \textit{Right panel}: Mean surface brightness within the effective radius of Coma cluster galaxies versus their $V$-band apparent magnitudes. The dashed lines correspond to constant effective radius. The outlines and the black crosses are the same as shown in the left panel.}
\end{figure*}

\begin{figure*}
    \includegraphics[width=0.96\textwidth]{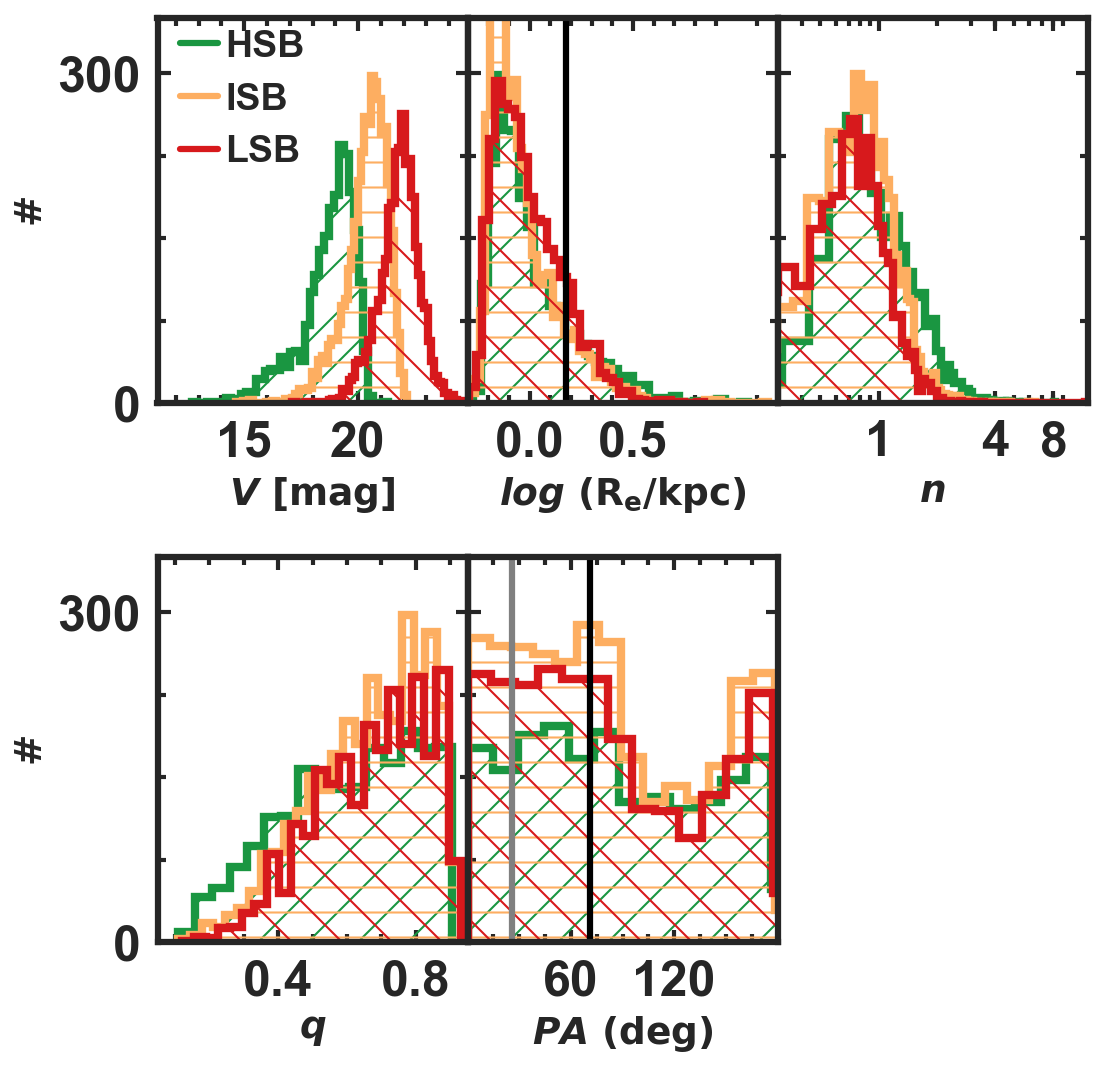}\hspace{0.01\textwidth}
	\caption{\label{fig:summ_hist} Distribution of the structural parameters from $V$-band  \texttt{GALFIT} analysis of Coma cluster galaxies in bins of high (HSB), intermediate (ISB), and low (LSB) mean surface brightness, as defined in the text. \textit{Top panels} show the distributions of the $V$-band magnitudes, physical sizes, \Sersic/ indices while the \textit{bottom panels} show the axial ratios and the folded position angles of Coma cluster galaxies, respectively. Within the size and surface brightness limits of our catalogue, the most likely Coma cluster galaxy has $V\unsim21$~mag, $\re\unsim0.8$~kpc, $n\unsim1$, and $q\unsim0.8$. The solid vertical line in the second subpanel is the fiducial size limit for UDGs. There is no obvious discontinuity at this size limit that would mark UDGs as a distinct galaxy subpopulation. In the position angle subpanel, we have highlighted the major-axis of the cluster with the black line and also show the $PA$ of NGC~4874, the central cD galaxy, with the gray line. We note that there is a remarkable drop in the number of galaxies with $PA\unsim120\degr$ in all surface brightness bins.}
\end{figure*}

\section{Summary of Results}
\label{sec:summ}
From our analysis thus far, we have obtained structural parameters (total magnitudes, $\re$, $n$, $q$, and $PA$ in both $V$ and $R$-bands) in $9,179$ galaxies in the direction of the Coma cluster. We show an example in Table~\ref{tab:complete_tab} and make the complete catalogue available online. Due to the depth of our imaging data, we also identified $233$ galaxies with faint, tidal features such as plumes, shells, rings, etc (see Figure~\ref{fig:NGC4911}). These features are the telltale signatures of the hierarchical nature of structure formation in the Universe. Table~\ref{tab:complete_tab} contains a summary of galaxies with such tidal features. 

As an example to highlight the wide range of galaxies in our catalogue, we show a $120$~kpc~$\times$~$120$~kpc field near NGC~4911 in Figure~\ref{fig:NGC4911}. This field contains galaxies ranging from the giant spiral galaxy (NGC~4911) to the ultra-diffuse galaxy (Y48), as well as several dwarf galaxies. In addition, the field also contains a disrupting galaxy ($\rm COMA\_11\_4120$), which as far as we know has not been previously catalogued in the literature. 

Figure~\ref{fig:summ_cat} shows the size--magnitude distribution of the Coma cluster galaxies. As expected, galaxy sizes correlate with their luminosities. In the $V$-band, we detect galaxies with mean surface brightness within the effective radius, $\langle \mu_{\rm eff,V} \rangle$, ranging from $20$ to $27.5$~\SBunit, effective radius, $\re$, from $\unsim0.6$ to $\unsim15$~kpc, and apparent magnitudes from $12$ to $24$~mag. We therefore span a region of parameter space that contains a diversity of galaxies suitable for a systematic exploration of the Coma cluster. As shown in both panels of Figure~\ref{fig:summ_cat}, our catalogue contains a large sample of low surface brightness galaxies ($\langle \mu_{\rm eff,V} \rangle \geq 24$~\SBunit) with small sizes ($\re \leq 1$~kpc) structurally similar to some Local Group dwarf galaxies. Our catalogue also contains galaxies with intermediate surface brightness $22 \leq \langle \mu_{\rm eff,V} \rangle \leq 24.5$ and $\re \geq 2.5$~kpc, a region of the parameter space relatively unexplored and often misunderstood as being devoid of galaxies (compare our Figure~\ref{fig:summ_cat} with fig.~$4$ from \citealt{Koda_2015}). This confirms the previous result from the \citet{Danieli_2019} where they also reported that this region is filled although their sample is incomplete below $\unsim2$~kpc.

We split our final catalogue into high surface brightness (HSB), intermediate surface brightness (ISB), and LSB galaxies using the following mean surface brightness limits -- HSB: $\langle \mu_{\rm eff,V} \rangle < 23$~\SBunit; ISB: $23 \geq \langle \mu_{\rm eff,V} \rangle < 24.5$~\SBunit; and LSB: $\langle \mu_{\rm eff,V} \rangle \geq 24.5$~\SBunit. These limits, which are in line with the categorization scheme used in \citet{Martin_2019}, are simply for convenience and we do not ascribe any particular astrophysical meaning to them. The bright limit of the LSB category is consistent with the LSB limit used in \protect\hyperlink{Y16}{Y16} catalogue. Our catalogue contains $2,290$~HSB, $3,833$~ISB, and $3,056$~LSB galaxies as defined above. This implies a factor of $3$ increase in the number of LSB galaxies in the Coma cluster relative to the \protect\hyperlink{Y16}{Y16} catalogue after accounting for background galaxy contamination. We show the distributions of the structural parameters in the $3$ surface brightness categories in Figure~\ref{fig:summ_hist} and note that regardless of surface brightness, the most likely galaxy in our catalogue has $V$-band magnitude~$\unsim21$~mag, $\re\unsim0.8$~kpc, $n\unsim1$, $q\unsim0.8$, and is most likely aligned along the $PA$ of the cluster, i.e. $\unsim71$~deg \citep{Plionis_1994}. The second subpanel shows that there is no obvious discontinuity in the sizes of the LSB galaxy subpopulation at $1.5$~kpc that would mark UDGs as a distinct galaxy population. 

Lastly, we note that \citet{Forbes_2020} used $V-R$ colours from this work, which were based on the difference between the total $V$ and total $R$ magnitudes, to explore the possible trends of globular cluster specific frequency with host UDG colour. Here, we have we used a different approach to calculate the $V-R$ colours of UDGs using colours from matched apertures as discussed in Section~\ref{sec:cmd}. This results in a similar specific frequency vs UDG colour trend to that found by \citet{Forbes_2020}.

\begin{figure*}
    \includegraphics[width=0.96\textwidth]{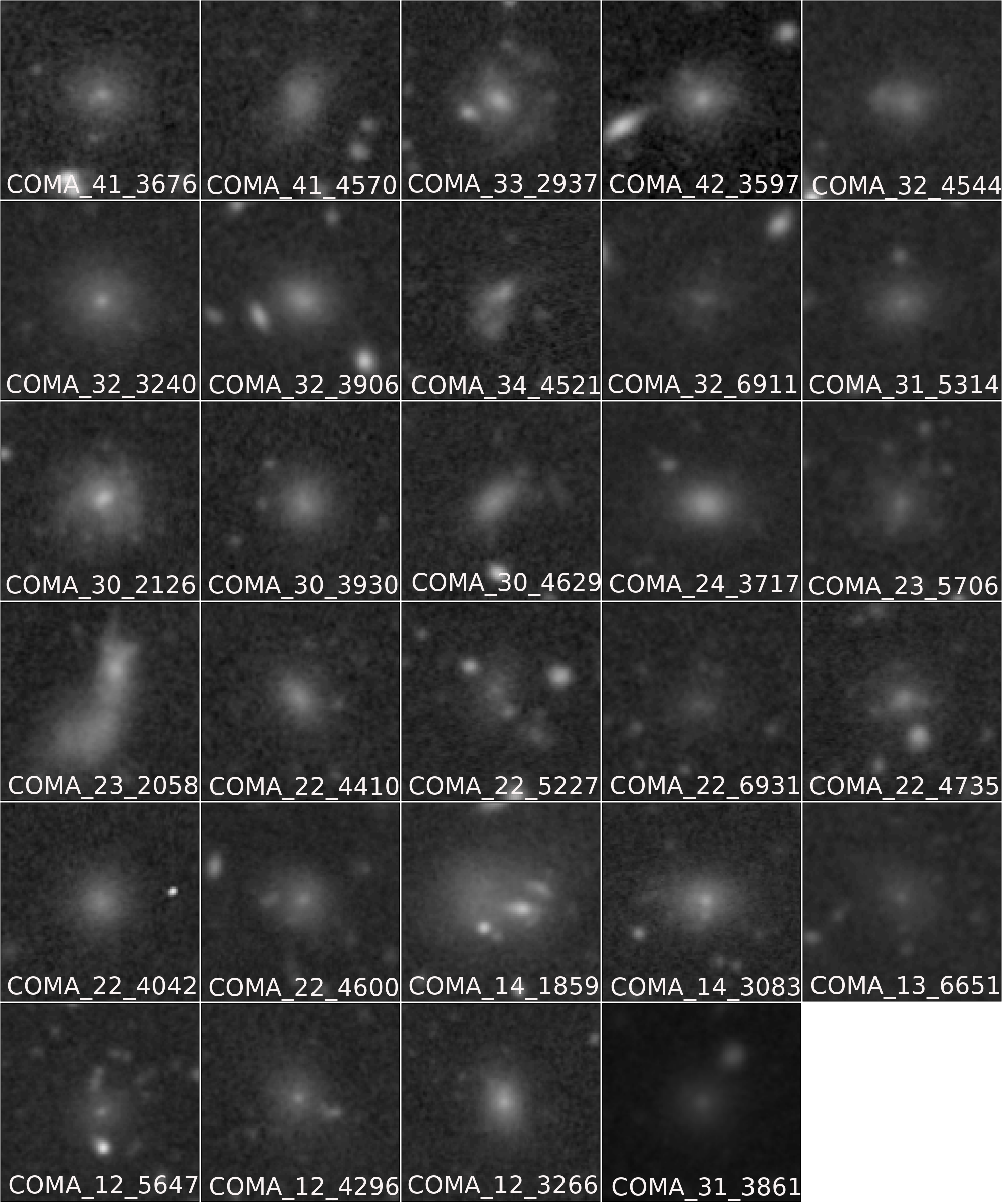}\hspace{0.01\textwidth}\\
	\caption{\label{fig:new_UDGs} $V$-band Subaru/Suprime-Cam thumbnails of the $29$ newly discovered Coma cluster ultra-diffuse galaxies. These UDGs have a typical $\re\unsim2$~kpc and $\langle \mu_{\rm eff,V} \rangle\unsim25$~\SBunit. Each thumbnail is $10\times10$~kpc across. North is up and East is to the left.}
\end{figure*}

\subsection{New ultra-diffuse galaxies in the Coma cluster}
\label{sec:new_udg}
Having established that our catalogue is rich in LSB galaxies, we immediately conduct a further search for new ultra-diffuse galaxies within the outlined UDG region shown in Figure~\ref{fig:summ_cat}. This region contains $110$ galaxies after excluding all the UDGs already identified in the literature, but only $29$ satisfy all the commonly used UDG criteria, i.e. $\re \geq 1.5$~kpc; mean surface brightness within $\re$,~$\langle \mu_{\rm eff,V} \rangle \geq24.5$~\SBunit\ (combining criterion 6 from \protect\hyperlink{Y16}{Y16} and using the mean UDG colour $V-R\unsim0.5$); and $\langle \mu_{\rm eff,V} \rangle - \mu_{\rm eff,V} \leq 0.8$ (see criterion 7 in \protect\hyperlink{Y16}{Y16}) where $\mu_{\rm eff,V}$ is the surface brightness at $\re$. We note that this last condition limits the sample of acceptable UDGs to those with \Sersic/ index, $n \leq 1.25$, i.e. exponential light profiles. Two factors may be responsible for these new detections: our data while covering the same sky area as that used in \protect\hyperlink{Y16}{Y16} had better seeing, and we have applied the detection methods originally introduced in \protect\hyperlink{Y16}{Y16} to multi-band data, making our measurements more reliable. We show these newly discovered UDGs in Figure~\ref{fig:new_UDGs} and present their structural parameters in Table~\ref{tab:UDGs}. Out of these newly catalogued UDGs, $23$ lie along the red-sequence of Coma cluster galaxies, i.e. they are RSG as discussed in Section~\ref{sec:cmd}, with the remaining $2$ having colours redder than the red-sequence region.

\begin{figure*}
    \includegraphics[width=0.96\textwidth]{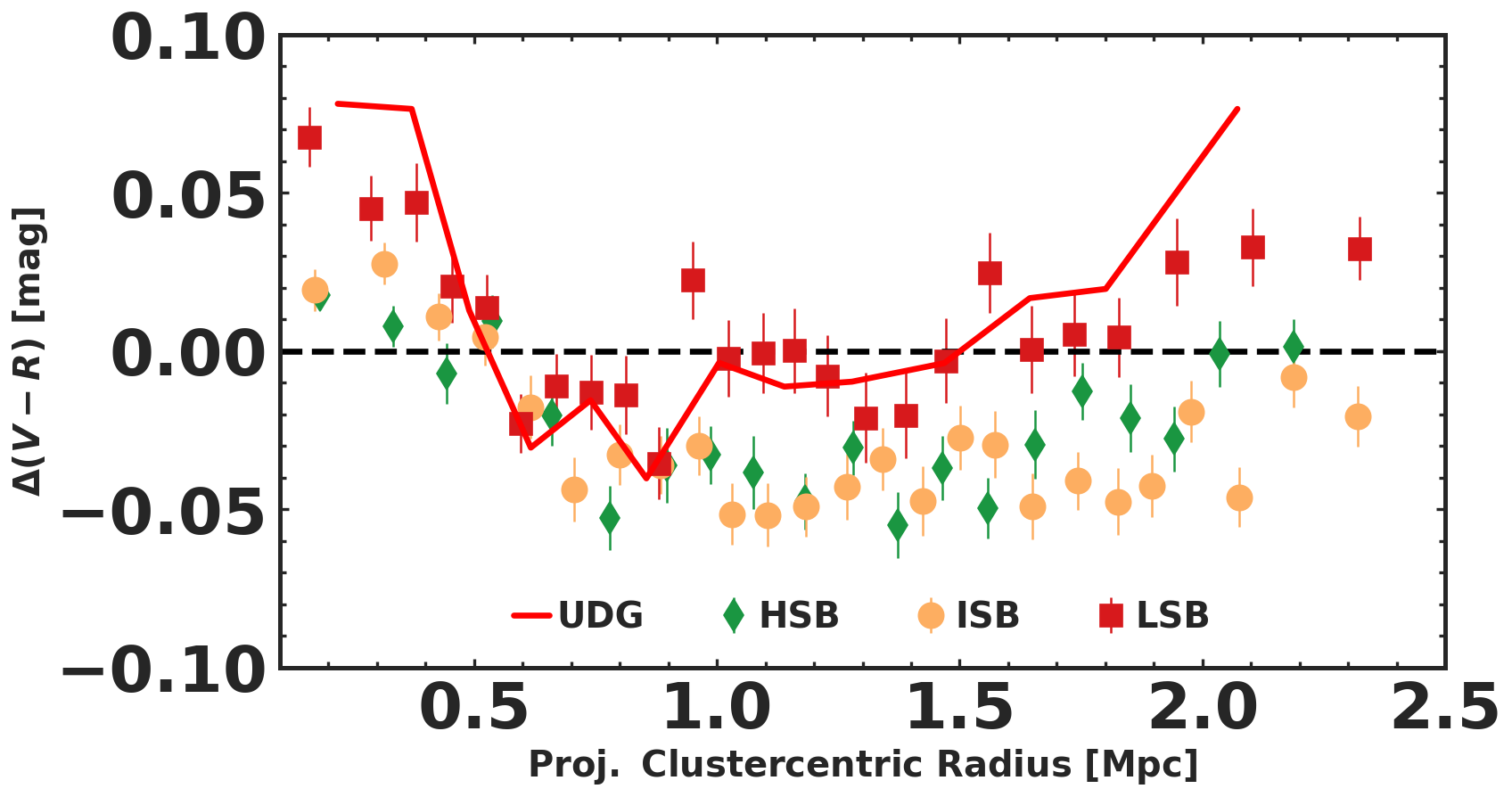}\hspace{0.01\textwidth}\\
	\caption{\label{fig:col_rad} Mean residual $V-R$ colours of Coma cluster galaxies after subtracting off the global red-sequence fit versus the projected clustercentric radius. We show results for the high (HSB), intermediate (ISB), and low (LSB) surface brightness galaxies, using the limits discussed in the text. Ultra-diffuse galaxies (UDG) which are represented by the solid red line have similar mean residual colour trends as the LSB galaxies out to the virial radius of the Coma cluster ($\unsim2.9$~Mpc). We do not include galaxies that are redder than the  $1\sigma$ scatter around the best-fit red-sequence line in this plot due to their high background contamination rate. The dashed horizontal line corresponds to galaxies with mean colours consistent with the red-sequence and it is shown to guide the eye. LSB galaxies within the cluster core have redder mean residual colours relative to the red-sequence relation at the $2\sigma$ level and they show a more dramatic transition at $\unsim0.6$~Mpc compared to galaxies with higher surface brightness, beyond which their mean residual colour flattens out, before rising in the cluster outskirts.}
\end{figure*}

\subsection{Environmental trends in clustercentric colour distribution}
\label{sec:rad_udg}
We now revisit the issue of environmental colour trends in the Coma cluster with emphasis on the LSB galaxies in our photometric catalogue. To recap, our catalogue contains $3,056$~LSB galaxies, most of which are along the red-sequence with $154$ and $631$ galaxies bluer and redder than the red-sequence region of the colour--magnitude diagram, respectively, as defined in Section~\ref{sec:cmd}. Our control field analysis in Section~\ref{sec:SDF} already shows that we can expect very little background galaxy contamination in the subsample that is bluer than the red-sequence, and up to $\unsim60$~per cent contamination in the LSB  galaxy subsample that is redder than the red-sequence. Along the red-sequence, the maximum background galaxy contamination level is $\unsim20$~per cent.

Galaxy clusters such as Coma are well known to show a decreasing colour gradient with projected clustercentric distance \citep{Terlevich_2001}, which is believed to be directly linked to a corresponding age gradient \citep{Smith_2009, Smith_2011}. The densest region of the cluster is observed to be dominated by redder galaxies while bluer galaxies dominate the less dense outskirt regions. Within the Coma cluster, the correlation of galaxy colours with their local environment has been previously observed in dwarf galaxies \citep{Secker_1997}, although \citet{Adami_2006} reported no radial trends in their sample of faint LSB galaxies. \citet{Alabi_2018}, however, found hints that UDGs within the cluster core may be redder than those in the cluster outskirts, although the analysis was severely limited by incomplete and inhomogeneously sourced colour data. More generally, LSB galaxies are expected to have formed their stars rapidly at earlier epochs \citep{Martin_2019} while those observed in dense environments may have experienced enhanced star-formation quenching via ram-pressure stripping and tidal puffing-up effects from their host clusters \citep{Moore_1996, Johansson_2009}. Imprints of such physical processes are expected to be seen in the radial distribution of the galaxy colours \citep{Jiang_2019}. \citet{Sales_2019} recently used the IllustrisTNG cosmological simulations to show that UDGs that were accreted early into present--day massive clusters ($\unsim10^{14}\Msun$) are found mostly within the cluster cores while those accreted more recently from the field environments dominate in the cluster outskirts.

Figure~\ref{fig:col_rad} shows how the mean residual $V-R$ colour distribution with respect to the best-fit red-sequence line varies with projected clustercentric radii out to the cluster virial radius, i.e. $\unsim2.9$~Mpc \citep{Kubo_2007}. We have excluded all galaxies that are redder than the $1\sigma$ scatter around the best-fit line of the red-sequence without spectroscopic confirmation as Coma cluster members in estimating the mean colour residuals. This is due to their high background galaxy contamination rate. Relative to the red-sequence, LSB galaxies (and UDGs) have redder mean colours compared to ISB and HSB galaxies within the cluster core. They have a mean residual colour that is $0.034\pm0.014$~mag redder compared to higher surface brightness galaxies. This result, which is significant at the $2\sigma$ level, suggests that LSB galaxies are most vulnerable to the severe star-formation quenching effects of the cluster-core environment. Outside the cluster-core, the mean residual colour trends are similar out to $\unsim1.8$~Mpc for the various surface brightness subsamples. The high mean residual colour at larger projected clustercentric radii, which is more evident in the LSB galaxies, can be attributed to an increase in the contribution from the redder, fainter background contamination galaxies to our final catalogue (see Section~\ref{sec:SDF}).

There is a noticeable transition at $\unsim0.6$~Mpc ($\unsim0.3R_{200}$), which is more pronounced in the LSB galaxies, already hinted at in \citet{Alabi_2018}. This transition radius corresponds to the projected radius within which UDGs that formed from cluster tidal effects dominate the UDG clustercentric number density profile in the cosmological simulations of \citet[][see their fig. 7]{Sales_2019}. The observed transition radius ($\unsim0.6$~Mpc) is similar to the the projected scale radius ($r_{\rm s}\unsim27\arcmin$ or $\unsim0.8$~Mpc at the distance of the Coma cluster) of the Coma cluster dark matter halo \citep{Okabe_2014}. Similar transition radii have been observed in the Virgo cluster \citep{Chung_2009} and more recently in galaxy clusters in the SAMI Galaxy Survey \citep{Owers_2019}. In both studies, the transition radius, although inferred from different parameters, was linked to environmental quenching effects which is most pronounced in the the cluster cores. For example, spiral galaxies within $\unsim0.5$~Mpc ($\unsim0.3R_{200}$) from the centre of Virgo cluster have HI disks that are smaller than their stellar disks, while in the SAMI Galaxy Survey, galaxies with strong Balmer absorption features but no recent star-formation episodes, were found exclusively within $\unsim0.6R_{200}$ of the cluster centres.

\section{Conclusion}
In this work, we have obtained structural parameters simultaneously in $V$ and $R$-bands for $9,179$ galaxies within an area of $\unsim4$~deg$^2$ in the Coma cluster using Subaru/Subprime-Cam data. Importantly, we have coherently obtained the $V-R$ colours of the ultra-diffuse galaxies in the Coma cluster, and in the more general class of low surface brightness galaxies, out to $\unsim2.6$~Mpc from the centre of the cluster. Our catalogue contains galaxies with magnitudes as faint as $V\unsim24$~mag, effective radius, $\re$, as small as $\unsim0.6$~kpc, and mean effective surface brightness within effective radius, $\langle \mu_{\rm eff,V} \rangle$, as low as $\unsim 27.5$~\SBunit. 

Our catalogue contains an unprecedented $3,056$ LSB galaxies in the direction of Coma cluster with mean effective surface brightness fainter than $24$~\SBunit~in the $V$-band. Out of these Coma cluster LSB galaxies, we found $29$ new UDGs, previously uncatalogued in the literature. In addition, we identified galaxies with faint tidal features within the Coma cluster. We make this catalogue publicly available. We have confirmed earlier results that most Coma cluster UDGs lie along the red-sequence of the colour--magnitude relation, but we found subpopulations of UDGs outside the red-sequence region.  

We also investigated clustercentric trends in galaxy colours in order to understand how the locally varying environment within the cluster affects the LSB galaxies compared to co-spatial higher surface brightness galaxies within the Coma cluster. We obtained the important result of a cluster transition radius at projected radius $\unsim0.6$~Mpc, within which the LSB galaxies are on average redder than galaxies with higher surface brightnesses at the $2\sigma$ level. This transition radius is similar to the region reported by \citet{Sales_2019} based on the IllustrisTNG cosmological simulation of massive clusters of galaxies within which ancient infalls dominate the UDG population.

\newpage
\section*{Acknowledgements}

We thank the anonymous referee for the thoughtful reading of the manuscript and for the valuable feedback.
DAF thanks the ARC for financial support via DP160101608. JPB and AJR acknowledge financial support from AST-1616598, AST-1518294 and 
AST-1616710. AJR was supported by the Research Corporation for Science Advancement as a Cottrell Scholar. This paper was based in part on data collected at the Subaru Telescope, which is operated by the National Astronomical Observatory of Japan. This research made use of the ``K-corrections calculator'' service available at http://kcor.sai.msu.ru/, \texttt{TOPCAT} software available at http://www.starlink.ac.uk/topcat/ \citep{Taylor_2005}, and \texttt{SEP} software \citep{Barbary_2016}.

\begin{sidewaystable*}
\scriptsize
\begin{tabular}{@{}l c c c c c c c c c c c c c c c l}
\hline
ID & RA & Dec & $V$  & $\re{_{_{V}}}$  &  $\langle \mu{_{\rm {eff},V}} \rangle$    & $n{_{_{V}}}$ & $q{_{_{V}}}$  & $PA{_{_{V}}}$ & $R$  & $\re{_{_{R}}}$  &  $\langle \mu{_{\rm {eff},R}} \rangle$    & $n{_{_{R}}}$ & $q{_{_{R}}}$  & $PA{_{_{R}}}$ &  $V-R$ & comment \\
     & [Degree] & [Degree]  & [mag] & [kpc] & [\SBunit]  &   &   & [deg] & [mag] & [kpc] & [\SBunit]  &   &   & [deg] & [mag] &                \\
      (1)  & (2)  &  (3)  & (4)  & (5)   & (6)   &  (7)  &  (8)  &  (9) &  (10) & (11)  & (12) & (13)  & (14) & (15) & (16) & (17)\\
\hline
  COMA\_10\_1339 & $195.41327 $ &  $27.160273 $ & $18.81\pm0.21$ & $4.31\pm 0.3 $  &  $25.55\pm 0.26$  &  $11.7 \pm50.62$  &  $0.89 \pm 0.14$ &  $80.79 $ &  $18.41\pm0.23$ & $5.87 \pm 0.4 $  &  $25.81 \pm 0.27$  &   $13.42\pm 0.69 $ & $0.88 \pm 0.14$  & $80.58 $ & $0.81$ &   $\rm LSB, >RS, GMP1928 $ \\
  COMA\_10\_2134 & $195.34398 $ &  $26.830603 $ & $19.36\pm0.18$ & $2.64\pm 0.22$  &  $25.04\pm 0.25$  &  $1.02 \pm 0.51$  &  $0.7  \pm 0.14$ &  $146.28$ &  $18.7 \pm0.16$ & $2.3  \pm 0.19$  &  $24.08 \pm 0.24$  &   $0.43 \pm 0.55 $ & $0.49 \pm 0.12$  & $141.41$ & $0.71$ &   $\rm LSB, Y16    $ \\ 
  COMA\_10\_2157 & $195.24165 $ &  $26.97631  $ & $19.37\pm0.22$ & $3.43\pm 0.24$  &  $25.62\pm 0.27$  &  $0.62 \pm 0.58$  &  $0.56 \pm 0.15$ &  $147.71$ &  $18.43\pm0.19$ & $3.74 \pm 0.23$  &  $24.86 \pm 0.23$  &   $0.66 \pm 0.59 $ & $0.6  \pm 0.13$  & $148.88$ & $0.71$ &   $\rm LSB, Y11, DF44     $ \\ 
  COMA\_10\_2222 & $195.31602 $ &  $27.210728 $ & $19.41\pm0.2 $ & $2.93\pm 0.23$  &  $25.31\pm 0.26$  &  $0.76 \pm 0.54$  &  $0.57 \pm 0.14$ &  $9.26  $ &  $18.58\pm0.17$ & $3.05 \pm 0.21$  &  $24.57 \pm 0.23$  &   $0.77 \pm 0.58 $ & $0.6  \pm 0.13$  & $9.08  $ & $0.62$ &   $\rm LSB, Y13,DFX1, GMP2175 $ \\
  COMA\_10\_3142 & $195.19983 $ &  $27.00997  $ & $19.9 \pm0.16$ & $1.77\pm 0.21$  &  $24.71\pm 0.3 $  &  $1.6  \pm 0.48$  &  $0.84 \pm 0.13$ &  $111.51$ &  $19.69\pm0.18$ & $1.91 \pm 0.22$  &  $24.66 \pm 0.3 $  &   $1.4  \pm 0.58 $ & $1.0  \pm 0.13$  & $20.13 $  & $0.38$ &  $\rm LSB      $ \\ 
  COMA\_10\_3630 & $195.32954 $ &  $27.054224 $ & $20.11\pm0.25$ & $2.92\pm 0.26$  &  $26.0 \pm 0.31$  &  $0.82 \pm 0.63$  &  $0.73 \pm 0.15$ &  $16.89 $ &  $19.69\pm0.21$ & $2.61 \pm 0.25$  &  $25.35 \pm 0.3 $  &   $0.82 \pm 0.61 $ & $0.87 \pm 0.13$  & $178.5 $ & $0.73$  &  $\rm LSB, Y14, DF42    $ \\ 
  COMA\_10\_4551 & $195.27219 $ &  $27.15981  $ & $20.51\pm0.17$ & $1.42\pm 0.21$  &  $24.85\pm 0.37$  &  $0.78 \pm 0.49$  &  $0.8  \pm 0.13$ &  $149.52$ &  $19.75\pm0.16$ & $1.43 \pm 0.19$  &  $24.1  \pm 0.33$  &   $0.81 \pm 0.56 $ & $0.8  \pm 0.12$  & $145.8 $ & $0.6$  &   $\rm LSB, Y12  $ \\ 
  COMA\_10\_4682 & $195.44453 $ &  $27.1667   $ & $20.57\pm0.18$ & $1.51\pm 0.22$  &  $25.03\pm 0.36$  &  $1.01 \pm 0.51$  &  $0.97 \pm 0.14$ &  $0.6   $ &  $19.81\pm0.17$ & $1.52 \pm 0.2 $  &  $24.29 \pm 0.33$  &   $0.98 \pm 0.56 $ & $0.98 \pm 0.13$  & $53.72 $ & $0.55$  &   $\rm LSB, Y18    $ \\ 
  COMA\_10\_5089 & $195.197   $ &  $26.78314  $ & $20.73\pm0.27$ & $2.48\pm 0.27$  &  $26.26\pm 0.36$  &  $0.73 \pm 0.67$  &  $0.66 \pm 0.16$ &  $141.95$ &  $20.07\pm0.22$ & $2.39 \pm 0.27$  &  $25.53 \pm 0.32$  &   $0.71 \pm 0.62 $ & $0.64 \pm 0.14$  & $140.62$ & $0.47$  &   $\rm LSB, Y8,DF46     $ \\ 
  COMA\_10\_5457 & $195.18846 $ &  $27.113832 $ & $20.87\pm0.2 $ & $1.53\pm 0.23$  &  $25.36\pm 0.39$  &  $0.99 \pm 0.55$  &  $0.89 \pm 0.14$ &  $112.1 $ &  $20.2 \pm0.18$ & $1.47 \pm 0.21$  &  $24.6  \pm 0.36$  &   $0.9  \pm 0.58 $ & $0.83 \pm 0.13$  & $114.37$ & $0.46$  &   $\rm LSB, Y6     $ \\ 
  COMA\_10\_5633 & $195.5619  $ &  $27.14568  $ & $20.94\pm0.32$ & $2.75\pm 0.3 $  &  $26.7 \pm 0.4 $  &  $1.6  \pm 0.74$  &  $0.69 \pm 0.17$ &  $54.99 $ &  $20.25\pm0.22$ & $2.39 \pm 0.28$  &  $25.71 \pm 0.34$  &   $1.21 \pm 0.63 $ & $0.77 \pm 0.14$  & $59.63 $ & $0.58$  &   $\rm LSB, Y22   $ \\ 
  COMA\_10\_5695 & $195.40529 $ &  $27.006954 $ & $20.97\pm0.18$ & $1.27\pm 0.22$  &  $25.06\pm 0.42$  &  $1.84 \pm 0.52$  &  $0.79 \pm 0.14$ &  $21.01 $ &  $20.96\pm0.2 $ & $1.29 \pm 0.24$  &  $25.08 \pm 0.45$  &   $1.43 \pm 0.6  $ & $0.82 \pm 0.13$  & $25.41 $ & $0.38$  &   $\rm LSB       $ \\ 
  COMA\_10\_5842 & $195.53276 $ &  $26.87714  $ & $21.03\pm0.16$ & $1.07\pm 0.21$  &  $24.75\pm 0.46$  &  $1.0  \pm 0.48$  &  $0.84 \pm 0.13$ &  $70.96 $ &  $20.93\pm0.18$ & $1.04 \pm 0.21$  &  $24.59 \pm 0.48$  &   $0.99 \pm 0.58 $ & $0.85 \pm 0.13$  & $70.04 $ & $0.41$  &   $\rm LSB        $ \\ 
  COMA\_10\_5902 & $195.27608 $ &  $27.168835 $ & $21.05\pm0.16$ & $1.04\pm 0.21$  &  $24.7 \pm 0.47$  &  $1.56 \pm 0.47$  &  $0.96 \pm 0.13$ &  $44.88 $ &  $20.82\pm0.17$ & $0.99 \pm 0.2 $  &  $24.37 \pm 0.47$  &   $1.52 \pm 0.57 $ & $0.73 \pm 0.13$  & $102.32$ & $0.51$  &   $\rm LSB        $ \\ 
  COMA\_10\_6090 & $195.30705 $ &  $26.866377 $ & $21.12\pm0.15$ & $0.92\pm 0.2 $  &  $24.51\pm 0.5 $  &  $0.27 \pm 0.46$  &  $0.52 \pm 0.13$ &  $123.13$ &  $20.33\pm0.14$ & $0.9  \pm 0.17$  &  $23.66 \pm 0.43$  &   $0.61 \pm 0.54 $ & $0.89 \pm 0.12$  & $101.31$  & $0.94$  &   $\rm LSB, >RS       $ \\ 
  COMA\_10\_6392 & $195.49966 $ &  $26.873947 $ & $21.24\pm0.16$ & $0.94\pm 0.21$  &  $24.67\pm 0.51$  &  $0.89 \pm 0.47$  &  $0.87 \pm 0.13$ &  $8.58  $ &  $20.89\pm0.17$ & $0.94 \pm 0.2 $  &  $24.31 \pm 0.49$  &   $0.9  \pm 0.56 $ & $0.92 \pm 0.13$  & $39.74 $ & $0.59$  &   $\rm LSB        $ \\ 
  COMA\_10\_6422 & $195.42538 $ &  $26.661331 $ & $21.25\pm0.15$ & $0.88\pm 0.2 $  &  $24.55\pm 0.53$  &  $1.03 \pm 0.46$  &  $0.79 \pm 0.13$ &  $3.59  $ &  $21.1 \pm0.17$ & $0.83 \pm 0.2 $  &  $24.27 \pm 0.54$  &   $1.06 \pm 0.56 $ & $0.81 \pm 0.13$  & $18.33 $ & $0.41$  &   $\rm LSB        $ \\ 
  COMA\_10\_6462 & $195.58606 $ &  $27.102407 $ & $21.26\pm0.25$ & $1.7 \pm 0.26$  &  $25.98\pm 0.41$  &  $0.79 \pm 0.63$  &  $0.46 \pm 0.15$ &  $14.57 $ &  $20.42\pm0.2 $ & $1.78 \pm 0.25$  &  $25.24 \pm 0.36$  &   $0.76 \pm 0.61 $ & $0.5  \pm 0.13$  & $11.53 $ & $0.62$  &   $\rm LSB, Y24   $ \\ 
  COMA\_10\_6536 & $195.48225 $ &  $27.10293  $ & $21.29\pm0.15$ & $0.86\pm 0.2 $  &  $24.54\pm 0.54$  &  $0.47 \pm 0.46$  &  $0.8  \pm 0.13$ &  $174.8 $ &  $20.93\pm0.16$ & $0.85 \pm 0.19$  &  $24.15 \pm 0.51$  &   $0.72 \pm 0.56 $ & $0.89 \pm 0.13$  & $5.25  $ & $0.59$  &   $\rm LSB       $ \\ 
  COMA\_10\_6577 & $195.130359$ &  $26.793341 $ & $21.31\pm0.16$ & $0.96\pm 0.21$  &  $24.77\pm 0.51$  &  $0.79 \pm 0.48$  &  $0.43 \pm 0.13$ &  $20.89 $ &  $20.76\pm0.16$ & $0.93 \pm 0.19$  &  $24.16 \pm 0.47$  &   $1.02 \pm 0.56 $ & $0.45 \pm 0.13$  & $17.05 $ & $0.8$  &   $\rm LSB, >RS        $ \\ 
  COMA\_10\_6622 & $195.5833  $ &  $27.209324 $ & $21.32\pm0.23$ & $1.52\pm 0.25$  &  $25.8 \pm 0.43$  &  $0.58 \pm 0.61$  &  $0.72 \pm 0.15$ &  $57.73 $ &  $20.57\pm0.19$ & $1.48 \pm 0.23$  &  $24.99 \pm 0.39$  &   $0.57 \pm 0.6  $ & $0.75 \pm 0.13$  & $55.81 $ & $0.6$  &   $\rm LSB, Y25     $ \\ 
  COMA\_10\_6670 & $195.3495  $ &  $26.75686  $ & $21.34\pm0.23$ & $1.47\pm 0.25$  &  $25.73\pm 0.43$  &  $0.83 \pm 0.6 $  &  $0.67 \pm 0.15$ &  $28.37 $ &  $20.62\pm0.19$ & $1.47 \pm 0.24$  &  $25.02 \pm 0.4 $  &   $0.83 \pm 0.6  $ & $0.73 \pm 0.13$  & $27.93 $ & $0.47$  &   $\rm LSB, Y15    $ \\ 
  COMA\_10\_6731 & $195.36357 $ &  $26.667072 $ & $21.38\pm0.22$ & $1.41\pm 0.25$  &  $25.69\pm 0.44$  &  $0.38 \pm 0.59$  &  $0.58 \pm 0.15$ &  $26.1  $ &  $21.24\pm0.21$ & $1.3  \pm 0.26$  &  $25.38 \pm 0.48$  &   $0.45 \pm 0.62 $ & $0.67 \pm 0.13$  & $29.89 $ & $0.49$  &   $\rm LSB        $ \\ 
  COMA\_10\_6914 & $195.12947 $ &  $27.020304 $ & $21.45\pm0.29$ & $1.93\pm 0.28$  &  $26.45\pm 0.43$  &  $0.64 \pm 0.7 $  &  $0.75 \pm 0.16$ &  $155.83$ &  $20.73\pm0.23$ & $2.0  \pm 0.28$  &  $25.8  \pm 0.38$  &   $0.81 \pm 0.64 $ & $1.0  \pm 0.14$  & $152.38$ & $0.58$  &   $\rm LSB, Y3    $ \\ 
  COMA\_10\_7073 & $195.24005 $ &  $26.781855 $ & $21.52\pm0.19$ & $1.1 \pm 0.23$  &  $25.29\pm 0.49$  &  $0.5  \pm 0.54$  &  $0.73 \pm 0.14$ &  $80.52 $ &  $21.2 \pm0.18$ & $1.0  \pm 0.22$  &  $24.76 \pm 0.51$  &   $0.68 \pm 0.59 $ & $0.76 \pm 0.13$  & $75.55 $ & $0.53$  &   $\rm LSB        $ \\ 
  COMA\_10\_7100 & $195.54    $ &  $27.135769 $ & $21.53\pm0.19$ & $1.04\pm 0.23$  &  $25.2 \pm 0.51$  &  $0.82 \pm 0.53$  &  $0.66 \pm 0.14$ &  $60.87 $ &  $21.26\pm0.19$ & $1.04 \pm 0.23$  &  $24.91 \pm 0.51$  &   $0.9  \pm 0.59 $ & $0.62 \pm 0.13$  & $58.82 $ & $0.53$  &   $\rm LSB      $ \\ 
  COMA\_10\_7154 & $195.144787$ &  $27.16266  $ & $21.55\pm0.17$ & $0.91\pm 0.22$  &  $24.9 \pm 0.54$  &  $0.88 \pm 0.5 $  &  $0.89 \pm 0.14$ &  $146.76$ &  $21.42\pm0.18$ & $0.83 \pm 0.21$  &  $24.57 \pm 0.58$  &   $1.02 \pm 0.58 $ & $0.66 \pm 0.13$  & $113.28$ & $0.41$  &   $\rm LSB      $ \\ 
  COMA\_10\_7241 & $195.57613 $ &  $26.69556  $ & $21.59\pm0.27$ & $1.62\pm 0.27$  &  $26.21\pm 0.45$  &  $0.22 \pm 0.67$  &  $0.69 \pm 0.16$ &  $64.54 $ &  $20.97\pm0.22$ & $1.68 \pm 0.27$  &  $25.67 \pm 0.42$  &   $0.17 \pm 0.63 $ & $0.75 \pm 0.14$  & $65.6  $ & $0.4$  &   $\rm LSB, Y23  $ \\ 
  COMA\_10\_7270 & $195.244732$ &  $27.152234 $ & $21.61\pm0.16$ & $0.8 \pm 0.21$  &  $24.68\pm 0.59$  &  $0.3  \pm 0.47$  &  $0.64 \pm 0.13$ &  $76.16 $ &  $20.98\pm0.15$ & $0.76 \pm 0.18$  &  $23.95 \pm 0.54$  &   $0.38 \pm 0.55 $ & $0.66 \pm 0.12$  & $75.74 $ & $0.93$  &   $\rm LSB, >RS $ \\ 
  COMA\_10\_7279 & $195.54225 $ &  $26.934872 $ & $21.61\pm0.19$ & $1.02\pm 0.23$  &  $25.23\pm 0.52$  &  $0.95 \pm 0.54$  &  $0.91 \pm 0.14$ &  $39.0  $ &  $21.22\pm0.19$ & $1.1  \pm 0.23$  &  $25.0  \pm 0.5 $  &   $1.05 \pm 0.6  $ & $0.93 \pm 0.13$  & $54.9  $ & $0.63$ &  $\rm LSB, >RS $ \\ 
  COMA\_10\_7296 & $195.25078 $ &  $26.677841 $ & $21.62\pm0.17$ & $0.9 \pm 0.22$  &  $24.96\pm 0.56$  &  $1.17 \pm 0.5 $  &  $0.53 \pm 0.14$ &  $6.42  $ &  $20.97\pm0.16$ & $0.86 \pm 0.19$  &  $24.22 \pm 0.52$  &   $1.4  \pm 0.56 $ & $0.53 \pm 0.13$  & $7.34  $ & $0.85$  &   $\rm LSB, >RS $ \\ 
  COMA\_10\_7323 & $195.48868 $ &  $26.91578  $ & $21.63\pm0.15$ & $0.76\pm 0.2 $  &  $24.59\pm 0.61$  &  $0.96 \pm 0.46$  &  $0.85 \pm 0.13$ &  $57.95 $ &  $21.24\pm0.16$ & $0.7  \pm 0.19$  &  $24.03 \pm 0.6 $  &   $1.0  \pm 0.55 $ & $0.87 \pm 0.12$  & $59.36 $ & $0.62$  &   $\rm LSB     $ \\ 
  COMA\_10\_7374 & $195.10555 $ &  $27.184435 $ & $21.65\pm0.18$ & $0.92\pm 0.22$  &  $25.03\pm 0.55$  &  $0.62 \pm 0.51$  &  $0.5  \pm 0.14$ &  $35.87 $ &  $21.02\pm0.17$ & $0.93 \pm 0.2 $  &  $24.44 \pm 0.51$  &   $0.75 \pm 0.57 $ & $0.54 \pm 0.13$  & $38.77 $ & $0.5$  &   $\rm LSB, Y2  $ \\ 
  COMA\_10\_7441 & $195.551728$ &  $27.151625 $ & $21.69\pm0.15$ & $0.74\pm 0.21$  &  $24.62\pm 0.62$  &  $0.39 \pm 0.47$  &  $0.52 \pm 0.13$ &  $5.03  $ &  $21.12\pm0.15$ & $0.68 \pm 0.18$  &  $23.86 \pm 0.59$  &   $0.56 \pm 0.55 $ & $0.46 \pm 0.12$  & $178.49$ & $0.87$ &   $\rm LSB, >RS   $ \\ 
  COMA\_10\_7508 & $195.283688$ &  $27.076344 $ & $21.72\pm0.15$ & $0.73\pm 0.21$  &  $24.6 \pm 0.63$  &  $0.1  \pm 0.46$  &  $0.74 \pm 0.13$ &  $48.81 $ &  $21.43\pm0.16$ & $0.71 \pm 0.19$  &  $24.24 \pm 0.62$  &   $0.06 \pm 0.56 $ & $0.85 \pm 0.13$  & $58.59 $ & $0.57$  &  $\rm LSB  $ \\ 
  COMA\_10\_7564 & $195.32216 $ &  $26.761515 $ & $21.74\pm0.2 $ & $1.01\pm 0.23$  &  $25.34\pm 0.54$  &  $1.02 \pm 0.55$  &  $0.57 \pm 0.14$ &  $159.69$ &  $21.73\pm0.19$ & $0.82 \pm 0.23$  &  $24.86 \pm 0.63$  &   $1.11 \pm 0.59 $ & $0.67 \pm 0.13$  & $140.72$ & $0.52$  &   $\rm LSB $ \\ 
  COMA\_10\_7634 & $195.30801 $ &  $26.792051 $ & $21.77\pm0.15$ & $0.69\pm 0.2 $  &  $24.54\pm 0.66$  &  $1.37 \pm 0.46$  &  $0.73 \pm 0.13$ &  $171.36$ &  $21.39\pm0.16$ & $0.65 \pm 0.19$  &  $24.03 \pm 0.64$  &   $1.06 \pm 0.55 $ & $0.79 \pm 0.12$  & $161.72$ & $0.68$  &   $\rm LSB, >RS $ \\ 

$\cdots$  & $\cdots$  &	$\cdots$ &	$\cdots$ &	$\cdots$   &$\cdots$ &	$\cdots$ &	$\cdots$ &	$\cdots$ &	$\cdots$ & $\cdots$  &	$\cdots$ & $\cdots$ & $\cdots$  &	$\cdots$ & $\cdots$\\
\hline   
\hline
\end{tabular}
\caption{Catalogue of Coma cluster galaxies with structural parameters from \texttt{GALFIT} analysis. This is an example page from the complete catalogue which is available online. The columns are (1): Galaxy ID, written to indicate the corresponding field, (2) and (3): Galaxy position (RA and Dec, respectively) in degrees (J2000), (4)--(9): and (10)--(15): are the total AB magnitude, circularized effective radius, mean surface brightness within the effective radius, \Sersic/ index, axial ratio (all with corresponding uncertainties derived following the description in the text), and position angle from $V$ and $R$-band analysis, respectively. We have applied K-correction and Galactic extinction correction to our photometry. $PA$ is measured East from North, (16): Galaxy colour obtained from \texttt{SExtractor} analysis, (17): comments about galaxy, starting with surface brightness category as defined in Section~\ref{sec:summ} (HSB, ISB or LSB), galaxies with colours redder than the 1$\sigma$ intrinsic scatter around the red-sequence are identified with $>$RS as explained in the text, identifier from the literature -- Identifier from \protect\hyperlink{Y16}{Y16}, Identifier from \citet{GMP_1983} and tidal features -- (I: interacting; D: disturbed morphology; JF: Jelly-fish; R: Rings; P: Plumes; S: Shells)}
\label{tab:complete_tab} 
\end{sidewaystable*}

\begin{table*}
\begin{tabular}{@{}l c c c c c c c c c}
\hline
ID & RA & Dec & $V$  & $\re{_{_{V}}}$  &  $\langle \mu{_{\rm {eff},V}} \rangle$    & $n{_{_{V}}}$ & $q{_{_{V}}}$  & $PA{_{_{V}}}$ & $V-R$\\
     & [Degree] & [Degree]  & [mag] & [kpc] & [\SBunit]  &   &   & [deg] & [mag] \\
      (1)  & (2)  &  (3)  & (4)  & (5)   & (6)   &  (7)  &  (8)  &  (9)  & (10) \\
\hline
  COMA\_14\_1859 & $195.21524$ & $28.871029$ & $19.21 \pm 0.16$  & $2.41 \pm 0.21$  &  $24.69 \pm 0.25$  &  $0.51 \pm 0.47$  &  $0.95 \pm 0.13$ &  $ 74$ &  $ 0.52$\\      
  COMA\_23\_2058 & $194.72876$ & $28.65372 $ & $19.32 \pm 0.16$  & $2.3  \pm 0.21$  &  $24.7  \pm 0.25$  &  $0.32 \pm 0.47$  &  $0.36 \pm 0.13$ &  $156$ &  $ 0.47$\\     
  COMA\_30\_2126 & $194.51813$ & $27.056221$ & $19.36 \pm 0.15$  & $2.1  \pm 0.2 $  &  $24.54 \pm 0.26$  &  $1.11 \pm 0.46$  &  $1.0  \pm 0.13$ &  $ 83$  & $0.46$\\     
  COMA\_33\_2937 & $194.4494 $ & $28.542898$ & $19.79 \pm 0.17$  & $2.03 \pm 0.22$  &  $24.9  \pm 0.29$  &  $1.24 \pm 0.5 $  &  $0.94 \pm 0.14$ &  $ 15$ & $0.44$\\     
  COMA\_14\_3083 & $195.15186$ & $28.930553$ & $19.87 \pm 0.17$  & $1.91 \pm 0.21$  &  $24.84 \pm 0.29$  &  $1.05 \pm 0.49$  &  $0.71 \pm 0.13$ &  $179$ & $0.47$\\     
  COMA\_32\_3240 & $194.50377$ & $27.88233 $ & $19.95 \pm 0.18$  & $2.08 \pm 0.22$  &  $25.11 \pm 0.3 $  &  $0.95 \pm 0.52$  &  $0.91 \pm 0.14$ &  $ 56$ & $0.39$\\     
  COMA\_12\_3266 & $195.17593$ & $28.164309$ & $19.96 \pm 0.17$  & $1.84 \pm 0.21$  &  $24.85 \pm 0.3 $  &  $1.21 \pm 0.49$  &  $0.84 \pm 0.13$ &  $ 21$ & $0.45$\\     
  COMA\_42\_3597 & $193.80339$ & $27.786472$ & $20.1  \pm 0.16$  & $1.62 \pm 0.21$  &  $24.71 \pm 0.32$  &  $1.01 \pm 0.48$  &  $0.76 \pm 0.13$ &  $ 92$ & $0.46$\\     
  COMA\_41\_3676 & $193.95773$ & $27.398582$ & $20.13 \pm 0.19$  & $2.04 \pm 0.23$  &  $25.25 \pm 0.31$  &  $1.13 \pm 0.54$  &  $0.99 \pm 0.14$ &  $ 80$ & $0.36$\\     
  COMA\_24\_3717 & $194.97047$ & $28.7967  $ & $20.15 \pm 0.15$  & $1.5  \pm 0.21$  &  $24.6  \pm 0.34$  &  $1.02 \pm 0.46$  &  $0.7  \pm 0.13$ &  $ 88$ & $0.5$\\     
  COMA\_31\_3861 & $194.49976$ & $27.619015$ & $20.21 \pm 0.17$  & $1.74 \pm 0.22$  &  $24.98 \pm 0.32$  &  $1.08 \pm 0.5 $  &  $0.99 \pm 0.14$ &  $120$ & $0.46$\\     
  COMA\_32\_3906 & $194.32854$ & $28.154984$ & $20.23 \pm 0.16$  & $1.53 \pm 0.21$  &  $24.72 \pm 0.34$  &  $1.0  \pm 0.48$  &  $0.81 \pm 0.13$ &  $ 63$ & $0.38$\\    
  COMA\_30\_3930 & $194.60199$ & $26.745077$ & $20.24 \pm 0.16$  & $1.58 \pm 0.21$  &  $24.81 \pm 0.34$  &  $0.86 \pm 0.49$  &  $0.9  \pm 0.13$ &  $ 12$ & $0.55$\\    
  COMA\_22\_4042 & $194.72652$ & $28.08718 $ & $20.29 \pm 0.17$  & $1.64 \pm 0.22$  &  $24.93 \pm 0.34$  &  $0.84 \pm 0.5 $  &  $0.92 \pm 0.14$ &  $157$ & $0.48$\\    
  COMA\_12\_4296 & $195.2982 $ & $28.145302$ & $20.41 \pm 0.22$  & $2.11 \pm 0.24$  &  $25.6  \pm 0.33$  &  $1.04 \pm 0.58$  &  $0.71 \pm 0.15$ &  $ 47$ & $0.46$\\    
  COMA\_22\_4410 & $195.06532$ & $28.029459$ & $20.45 \pm 0.17$  & $1.5  \pm 0.22$  &  $24.9  \pm 0.36$  &  $1.08 \pm 0.5 $  &  $0.69 \pm 0.14$ &  $ 37$ & $0.54$\\    
  COMA\_34\_4521 & $194.25638$ & $28.87275 $ & $20.5  \pm 0.18$  & $1.58 \pm 0.22$  &  $25.06 \pm 0.35$  &  $0.85 \pm 0.51$  &  $0.75 \pm 0.14$ &  $ 36$ & $1.18$\\    
  COMA\_32\_4544 & $194.46745$ & $28.105352$ & $20.51 \pm 0.18$  & $1.58 \pm 0.22$  &  $25.07 \pm 0.35$  &  $0.83 \pm 0.52$  &  $0.92 \pm 0.14$ &  $ 70$ & $0.33$\\    
  COMA\_41\_4570 & $193.86931$ & $27.684788$ & $20.52 \pm 0.18$  & $1.55 \pm 0.22$  &  $25.05 \pm 0.36$  &  $0.75 \pm 0.51$  &  $0.62 \pm 0.14$ &  $158$ & $0.41$\\    
  COMA\_22\_4600 & $194.71304$ & $28.13771 $ & $20.54 \pm 0.18$  & $1.58 \pm 0.22$  &  $25.1  \pm 0.36$  &  $0.63 \pm 0.52$  &  $0.84 \pm 0.14$ &  $ 72$ & $0.52$\\    
  COMA\_30\_4629 & $194.23543$ & $27.045115$ & $20.55 \pm 0.17$  & $1.48 \pm 0.22$  &  $24.96 \pm 0.36$  &  $0.76 \pm 0.5 $  &  $0.99 \pm 0.14$ &  $ 60$ & $0.64$\\    
  COMA\_22\_4735 & $194.77484$ & $27.97042 $ & $20.6  \pm 0.18$  & $1.56 \pm 0.22$  &  $25.13 \pm 0.36$  &  $0.29 \pm 0.52$  &  $0.46 \pm 0.14$ &  $ 21$ & $0.55$\\    
  COMA\_22\_5227 & $194.9666 $ & $27.82323 $ & $20.79 \pm 0.21$  & $1.71 \pm 0.24$  &  $25.52 \pm 0.37$  &  $0.44 \pm 0.57$  &  $0.33 \pm 0.14$ &  $ 42$ & $0.71$\\    
  COMA\_31\_5314 & $194.59314$ & $27.38581 $ & $20.82 \pm 0.2 $  & $1.53 \pm 0.23$  &  $25.31 \pm 0.38$  &  $0.78 \pm 0.54$  &  $0.76 \pm 0.14$ &  $100$ & $0.45$\\    
  COMA\_12\_5647 & $195.48859$ & $28.138306$ & $20.95 \pm 0.22$  & $1.73 \pm 0.25$  &  $25.71 \pm 0.38$  &  $1.03 \pm 0.6 $  &  $0.95 \pm 0.15$ &  $161$ & $0.53$\\    
  COMA\_23\_5706 & $194.8023 $ & $28.353645$ & $20.97 \pm 0.23$  & $1.8  \pm 0.25$  &  $25.82 \pm 0.38$  &  $1.06 \pm 0.61$  &  $0.99 \pm 0.15$ &  $ 39$ & $0.29$\\    
  COMA\_13\_6651 & $195.3371 $ & $28.539967$ & $21.34 \pm 0.25$  & $1.7  \pm 0.26$  &  $26.05 \pm 0.42$  &  $0.7  \pm 0.64$  &  $0.77 \pm 0.16$ &  $ 46$ & $0.4$\\    
  COMA\_32\_6911 & $194.60907$ & $27.867151$ & $21.45 \pm 0.25$  & $1.55 \pm 0.26$  &  $25.96 \pm 0.44$  &  $1.01 \pm 0.63$  &  $0.98 \pm 0.15$ &  $122$ & $0.28$\\    
  COMA\_22\_6931 & $194.78117$ & $28.13321 $ & $21.46 \pm 0.27$  & $1.73 \pm 0.27$  &  $26.21 \pm 0.43$  &  $1.02 \pm 0.67$  &  $0.81 \pm 0.16$ &  $133$ & $0.5$\\  

\hline   
\hline
\end{tabular}
\caption{Newly discovered ultra-diffuse galaxies in the Coma cluster catalogued here for convenience. Columns are the same as in Table~\ref{tab:complete_tab} although we only show parameters from the $V$-band.}
\label{tab:UDGs} 
\end{table*}

\section*{Data Availability}
The data underlying this article are available in the article and in its online supplementary material.

\bibliographystyle{mnras}
\bibliography{coma}

\begin{thebibliography}{}
\makeatletter
\relax
\def\mn@urlcharsother{\let\do\@makeother \do\$\do\&\do\#\do\^\do\_\do\%\do\~}
\def\mn@doi{\begingroup\mn@urlcharsother \@ifnextchar [ {\mn@doi@}
  {\mn@doi@[]}}
\def\mn@doi@[#1]#2{\def\@tempa{#1}\ifx\@tempa\@empty \href
  {http://dx.doi.org/#2} {doi:#2}\else \href {http://dx.doi.org/#2} {#1}\fi
  \endgroup}
\def\mn@eprint#1#2{\mn@eprint@#1:#2::\@nil}
\def\mn@eprint@arXiv#1{\href {http://arxiv.org/abs/#1} {{\tt arXiv:#1}}}
\def\mn@eprint@dblp#1{\href {http://dblp.uni-trier.de/rec/bibtex/#1.xml}
  {dblp:#1}}
\def\mn@eprint@#1:#2:#3:#4\@nil{\def\@tempa {#1}\def\@tempb {#2}\def\@tempc
  {#3}\ifx \@tempc \@empty \let \@tempc \@tempb \let \@tempb \@tempa \fi \ifx
  \@tempb \@empty \def\@tempb {arXiv}\fi \@ifundefined
  {mn@eprint@\@tempb}{\@tempb:\@tempc}{\expandafter \expandafter \csname
  mn@eprint@\@tempb\endcsname \expandafter{\@tempc}}}

\bibitem[\protect\citeauthoryear{{Abraham} \& {van Dokkum}}{{Abraham} \& {van
  Dokkum}}{2014}]{Abraham_2014}
{Abraham} R.~G.,  {van Dokkum} P.~G.,  2014, \mn@doi [\pasp] {10.1086/674875},
  \href {https://ui.adsabs.harvard.edu/abs/2014PASP..126...55A} {126, 55}

\bibitem[\protect\citeauthoryear{{Adami} et~al.,}{{Adami}
  et~al.}{2006}]{Adami_2006}
{Adami} C.,  et~al., 2006, \mn@doi [\aap] {10.1051/0004-6361:20053758}, \href
  {https://ui.adsabs.harvard.edu/abs/2006A&A...459..679A} {459, 679}

\bibitem[\protect\citeauthoryear{{Adami} et~al.,}{{Adami}
  et~al.}{2009}]{Adami_2009}
{Adami} C.,  et~al., 2009, \mn@doi [\aap] {10.1051/0004-6361/200912228}, \href
  {https://ui.adsabs.harvard.edu/abs/2009A&A...507.1225A} {507, 1225}

\bibitem[\protect\citeauthoryear{{Aguerri}, {Iglesias-P{\'a}ramo},
  {V{\'\i}lchez}, {Mu{\~n}oz-Tu{\~n}{\'o}n}  \&
  {S{\'a}nchez-Janssen}}{{Aguerri} et~al.}{2005}]{Aguerri_2005}
{Aguerri} J.~A.~L.,  {Iglesias-P{\'a}ramo} J.,  {V{\'\i}lchez} J.~M.,
  {Mu{\~n}oz-Tu{\~n}{\'o}n} C.,   {S{\'a}nchez-Janssen} R.,  2005, \mn@doi
  [\aj] {10.1086/431360}, \href
  {https://ui.adsabs.harvard.edu/abs/2005AJ....130..475A} {130, 475}

\bibitem[\protect\citeauthoryear{{Ahn} et~al.,}{{Ahn} et~al.}{2012}]{Ahn_2012}
{Ahn} C.~P.,  et~al., 2012, \mn@doi [\apjs] {10.1088/0067-0049/203/2/21}, \href
  {https://ui.adsabs.harvard.edu/abs/2012ApJS..203...21A} {203, 21}

\bibitem[\protect\citeauthoryear{{Alabi} et~al.,}{{Alabi}
  et~al.}{2018}]{Alabi_2018}
{Alabi} A.,  et~al., 2018, \mn@doi [\mnras] {10.1093/mnras/sty1616}, \href
  {https://ui.adsabs.harvard.edu/abs/2018MNRAS.479.3308A} {479, 3308}

\bibitem[\protect\citeauthoryear{{Barbary}}{{Barbary}}{2016}]{Barbary_2016}
{Barbary} K.,  2016, \mn@doi [The Journal of Open Source Software]
  {10.21105/joss.00058}, \href
  {https://ui.adsabs.harvard.edu/abs/2016JOSS....1...58B} {1, 58}

\bibitem[\protect\citeauthoryear{{Bertin} \& {Arnouts}}{{Bertin} \&
  {Arnouts}}{1996}]{Bertin_1996}
{Bertin} E.,  {Arnouts} S.,  1996, \mn@doi [\aaps] {10.1051/aas:1996164}, \href
  {https://ui.adsabs.harvard.edu/abs/1996A&AS..117..393B} {117, 393}

\bibitem[\protect\citeauthoryear{{Blanton} \& {Roweis}}{{Blanton} \&
  {Roweis}}{2007}]{Blanton_2007}
{Blanton} M.~R.,  {Roweis} S.,  2007, \mn@doi [\aj] {10.1086/510127}, \href
  {https://ui.adsabs.harvard.edu/abs/2007AJ....133..734B} {133, 734}

\bibitem[\protect\citeauthoryear{{Carter} et~al.,}{{Carter}
  et~al.}{2002}]{Carter_2002}
{Carter} D.,  et~al., 2002, \mn@doi [\apj] {10.1086/338667}, \href
  {https://ui.adsabs.harvard.edu/abs/2002ApJ...567..772C} {567, 772}

\bibitem[\protect\citeauthoryear{{Carter} et~al.,}{{Carter}
  et~al.}{2008}]{Carter_2008}
{Carter} D.,  et~al., 2008, \mn@doi [\apjs] {10.1086/533439}, \href
  {https://ui.adsabs.harvard.edu/abs/2008ApJS..176..424C} {176, 424}

\bibitem[\protect\citeauthoryear{{Chiboucas}, {Tully}, {Marzke}, {Trentham},
  {Ferguson}, {Hammer}, {Carter}  \& {Khosroshahi}}{{Chiboucas}
  et~al.}{2010}]{Chiboucas_2010}
{Chiboucas} K.,  {Tully} R.~B.,  {Marzke} R.~O.,  {Trentham} N.,  {Ferguson}
  H.~C.,  {Hammer} D.,  {Carter} D.,   {Khosroshahi} H.,  2010, \mn@doi [\apj]
  {10.1088/0004-637X/723/1/251}, \href
  {https://ui.adsabs.harvard.edu/abs/2010ApJ...723..251C} {723, 251}

\bibitem[\protect\citeauthoryear{{Chilingarian} \& {Zolotukhin}}{{Chilingarian}
  \& {Zolotukhin}}{2012}]{Chilingarian_2012}
{Chilingarian} I.~V.,  {Zolotukhin} I.~Y.,  2012, \mn@doi [\mnras]
  {10.1111/j.1365-2966.2011.19837.x}, \href
  {https://ui.adsabs.harvard.edu/abs/2012MNRAS.419.1727C} {419, 1727}

\bibitem[\protect\citeauthoryear{{Chilingarian}, {Melchior}  \&
  {Zolotukhin}}{{Chilingarian} et~al.}{2010}]{Chilingarian_2010}
{Chilingarian} I.~V.,  {Melchior} A.-L.,   {Zolotukhin} I.~Y.,  2010, \mn@doi
  [\mnras] {10.1111/j.1365-2966.2010.16506.x}, \href
  {https://ui.adsabs.harvard.edu/abs/2010MNRAS.405.1409C} {405, 1409}

\bibitem[\protect\citeauthoryear{Chilingarian, Afanasiev, Grishin, Fabricant
  \& Moran}{Chilingarian et~al.}{2019}]{Chilingarian_2019}
Chilingarian I.~V.,  Afanasiev A.~V.,  Grishin K.~A.,  Fabricant D.,   Moran
  S.,  2019, \mn@doi [The Astrophysical Journal] {10.3847/1538-4357/ab4205},
  884, 79

\bibitem[\protect\citeauthoryear{Chung, van Gorkom, Kenney, Crowl  \&
  Vollmer}{Chung et~al.}{2009}]{Chung_2009}
Chung A.,  van Gorkom J.~H.,  Kenney J. D.~P.,  Crowl H.,   Vollmer B.,  2009,
  \mn@doi [The Astronomical Journal] {10.1088/0004-6256/138/6/1741}, 138, 1741

\bibitem[\protect\citeauthoryear{{Danieli} \& {van Dokkum}}{{Danieli} \& {van
  Dokkum}}{2019}]{Danieli_2019}
{Danieli} S.,  {van Dokkum} P.,  2019, \mn@doi [\apj]
  {10.3847/1538-4357/ab14f3}, \href
  {https://ui.adsabs.harvard.edu/abs/2019ApJ...875..155D} {875, 155}

\bibitem[\protect\citeauthoryear{{Edwards}, {Colless}, {Bridges}, {Carter},
  {Mobasher}  \& {Poggianti}}{{Edwards} et~al.}{2002}]{Edwards_2002}
{Edwards} S.~A.,  {Colless} M.,  {Bridges} T.~J.,  {Carter} D.,  {Mobasher} B.,
    {Poggianti} B.~M.,  2002, \mn@doi [\apj] {10.1086/338495}, \href
  {https://ui.adsabs.harvard.edu/abs/2002ApJ...567..178E} {567, 178}

\bibitem[\protect\citeauthoryear{Eisenhardt, Propris, Gonzalez, Stanford, Wang
  \& Dickinson}{Eisenhardt et~al.}{2007}]{Eisenhardt_2007}
Eisenhardt P.~R.,  Propris R.~D.,  Gonzalez A.~H.,  Stanford S.~A.,  Wang M.,
  Dickinson M.,  2007, \mn@doi [The Astrophysical Journal Supplement Series]
  {10.1086/511688}, 169, 225

\bibitem[\protect\citeauthoryear{{Ferr{\'e}-Mateu} et~al.,}{{Ferr{\'e}-Mateu}
  et~al.}{2018}]{Ferre_2018}
{Ferr{\'e}-Mateu} A.,  et~al., 2018, \mn@doi [\mnras] {10.1093/mnras/sty1597},
  \href {https://ui.adsabs.harvard.edu/abs/2018MNRAS.479.4891F} {479, 4891}

\bibitem[\protect\citeauthoryear{{Forbes}, {Alabi}, {Romanowsky}, {Brodie}  \&
  {Arimoto}}{{Forbes} et~al.}{2020}]{Forbes_2020}
{Forbes} D.~A.,  {Alabi} A.,  {Romanowsky} A.~J.,  {Brodie} J.~P.,   {Arimoto}
  N.,  2020, \mn@doi [\mnras] {10.1093/mnras/staa180}, \href
  {https://ui.adsabs.harvard.edu/abs/2020MNRAS.492.4874F} {492, 4874}

\bibitem[\protect\citeauthoryear{{Godwin}, {Metcalfe}  \& {Peach}}{{Godwin}
  et~al.}{1983}]{GMP_1983}
{Godwin} J.~G.,  {Metcalfe} N.,   {Peach} J.~V.,  1983, \mn@doi [\mnras]
  {10.1093/mnras/202.1.113}, \href
  {https://ui.adsabs.harvard.edu/abs/1983MNRAS.202..113G} {202, 113}

\bibitem[\protect\citeauthoryear{{Graham} \& {Driver}}{{Graham} \&
  {Driver}}{2005}]{Graham_2005}
{Graham} A.~W.,  {Driver} S.~P.,  2005, \mn@doi [\pasa] {10.1071/AS05001},
  \href {https://ui.adsabs.harvard.edu/abs/2005PASA...22..118G} {22, 118}

\bibitem[\protect\citeauthoryear{{Guti{\'e}rrez}, {Trujillo}, {Aguerri},
  {Graham}  \& {Caon}}{{Guti{\'e}rrez} et~al.}{2004}]{Gutierrez_2004}
{Guti{\'e}rrez} C.~M.,  {Trujillo} I.,  {Aguerri} J. A.~L.,  {Graham} A.~W.,
  {Caon} N.,  2004, \mn@doi [\apj] {10.1086/381022}, \href
  {https://ui.adsabs.harvard.edu/abs/2004ApJ...602..664G} {602, 664}

\bibitem[\protect\citeauthoryear{{H{\"a}ussler} et~al.,}{{H{\"a}ussler}
  et~al.}{2007}]{Haussler_2007}
{H{\"a}ussler} B.,  et~al., 2007, \mn@doi [\apjs] {10.1086/518836}, \href
  {https://ui.adsabs.harvard.edu/abs/2007ApJS..172..615H} {172, 615}

\bibitem[\protect\citeauthoryear{{Hoyos} et~al.,}{{Hoyos}
  et~al.}{2011}]{Hoyos_2011}
{Hoyos} C.,  et~al., 2011, \mn@doi [\mnras] {10.1111/j.1365-2966.2010.17855.x},
  \href {https://ui.adsabs.harvard.edu/abs/2011MNRAS.411.2439H} {411, 2439}

\bibitem[\protect\citeauthoryear{{Jiang}, {Dekel}, {Freundlich}, {Romanowsky},
  {Dutton}, {Macci{\`o}}  \& {Di Cintio}}{{Jiang} et~al.}{2019}]{Jiang_2019}
{Jiang} F.,  {Dekel} A.,  {Freundlich} J.,  {Romanowsky} A.~J.,  {Dutton}
  A.~A.,  {Macci{\`o}} A.~V.,   {Di Cintio} A.,  2019, \mn@doi [\mnras]
  {10.1093/mnras/stz1499}, \href
  {https://ui.adsabs.harvard.edu/abs/2019MNRAS.487.5272J} {487, 5272}

\bibitem[\protect\citeauthoryear{{Johansson}, {Naab}  \&
  {Ostriker}}{{Johansson} et~al.}{2009}]{Johansson_2009}
{Johansson} P.~H.,  {Naab} T.,   {Ostriker} J.~P.,  2009, \mn@doi [\apjl]
  {10.1088/0004-637X/697/1/L38}, \href
  {https://ui.adsabs.harvard.edu/abs/2009ApJ...697L..38J} {697, L38}

\bibitem[\protect\citeauthoryear{{Kadowaki}, {Zaritsky}  \&
  {Donnerstein}}{{Kadowaki} et~al.}{2017}]{Kadowaki_2017}
{Kadowaki} J.,  {Zaritsky} D.,   {Donnerstein} R.~L.,  2017, \mn@doi [\apjl]
  {10.3847/2041-8213/aa653d}, \href
  {https://ui.adsabs.harvard.edu/abs/2017ApJ...838L..21K} {838, L21}

\bibitem[\protect\citeauthoryear{{Kashikawa} et~al.,}{{Kashikawa}
  et~al.}{2004}]{Kashikawa_2004}
{Kashikawa} N.,  et~al., 2004, \mn@doi [\pasj] {10.1093/pasj/56.6.1011}, \href
  {https://ui.adsabs.harvard.edu/abs/2004PASJ...56.1011K} {56, 1011}

\bibitem[\protect\citeauthoryear{{Kelly} et~al.,}{{Kelly}
  et~al.}{2014}]{Kelly_2014}
{Kelly} P.~L.,  et~al., 2014, \mn@doi [\mnras] {10.1093/mnras/stt1946}, \href
  {https://ui.adsabs.harvard.edu/abs/2014MNRAS.439...28K} {439, 28}

\bibitem[\protect\citeauthoryear{{Koda}, {Yagi}, {Yamanoi}  \&
  {Komiyama}}{{Koda} et~al.}{2015}]{Koda_2015}
{Koda} J.,  {Yagi} M.,  {Yamanoi} H.,   {Komiyama} Y.,  2015, \mn@doi [\apjl]
  {10.1088/2041-8205/807/1/L2}, \href
  {https://ui.adsabs.harvard.edu/abs/2015ApJ...807L...2K} {807, L2}

\bibitem[\protect\citeauthoryear{{Kron}}{{Kron}}{1980}]{Kron_1980}
{Kron} R.~G.,  1980, \mn@doi [\apjs] {10.1086/190669}, \href
  {https://ui.adsabs.harvard.edu/abs/1980ApJS...43..305K} {43, 305}

\bibitem[\protect\citeauthoryear{{Kubo}, {Stebbins}, {Annis}, {Dell'Antonio},
  {Lin}, {Khiabanian}  \& {Frieman}}{{Kubo} et~al.}{2007}]{Kubo_2007}
{Kubo} J.~M.,  {Stebbins} A.,  {Annis} J.,  {Dell'Antonio} I.~P.,  {Lin} H.,
  {Khiabanian} H.,   {Frieman} J.~A.,  2007, \mn@doi [\apj] {10.1086/523101},
  \href {http://adsabs.harvard.edu/abs/2007ApJ...671.1466K} {671, 1466}

\bibitem[\protect\citeauthoryear{{Lim}, {Peng}, {C{\^o}t{\'e}}, {Sales}, {den
  Brok}, {Blakeslee}  \& {Guhathakurta}}{{Lim} et~al.}{2018}]{Lim_2018}
{Lim} S.,  {Peng} E.~W.,  {C{\^o}t{\'e}} P.,  {Sales} L.~V.,  {den Brok} M.,
  {Blakeslee} J.~P.,   {Guhathakurta} P.,  2018, \mn@doi [\apj]
  {10.3847/1538-4357/aacb81}, \href
  {https://ui.adsabs.harvard.edu/abs/2018ApJ...862...82L} {862, 82}

\bibitem[\protect\citeauthoryear{{Mahajan}, {Haines}  \&
  {Raychaudhury}}{{Mahajan} et~al.}{2011}]{Mahajan_2011}
{Mahajan} S.,  {Haines} C.~P.,   {Raychaudhury} S.,  2011, \mn@doi [\mnras]
  {10.1111/j.1365-2966.2010.17977.x}, \href
  {https://ui.adsabs.harvard.edu/abs/2011MNRAS.412.1098M} {412, 1098}

\bibitem[\protect\citeauthoryear{Mancera~Piña, Aguerri, Peletier, Venhola,
  Trager  \& Choque~Challapa}{Mancera~Piña et~al.}{2019}]{Mancera_2019}
Mancera~Piña P.~E.,  Aguerri J. A.~L.,  Peletier R.~F.,  Venhola A.,  Trager
  S.,   Choque~Challapa N.,  2019, \mn@doi [Monthly Notices of the Royal
  Astronomical Society] {10.1093/mnras/stz238}, 485, 1036–1052

\bibitem[\protect\citeauthoryear{{Martin} et~al.,}{{Martin}
  et~al.}{2019}]{Martin_2019}
{Martin} G.,  et~al., 2019, \mn@doi [\mnras] {10.1093/mnras/stz356}, \href
  {https://ui.adsabs.harvard.edu/abs/2019MNRAS.485..796M} {485, 796}

\bibitem[\protect\citeauthoryear{{McConnachie}}{{McConnachie}}{2012}]{McConnachie_2012}
{McConnachie} A.~W.,  2012, \mn@doi [\aj] {10.1088/0004-6256/144/1/4}, \href
  {https://ui.adsabs.harvard.edu/abs/2012AJ....144....4M} {144, 4}

\bibitem[\protect\citeauthoryear{{Miyazaki} et~al.,}{{Miyazaki}
  et~al.}{2002}]{Miyazaki_2002}
{Miyazaki} S.,  et~al., 2002, \mn@doi [\pasj] {10.1093/pasj/54.6.833}, \href
  {https://ui.adsabs.harvard.edu/abs/2002PASJ...54..833M} {54, 833}

\bibitem[\protect\citeauthoryear{{Mobasher} et~al.,}{{Mobasher}
  et~al.}{2001}]{Mobasher_2001}
{Mobasher} B.,  et~al., 2001, \mn@doi [\apjs] {10.1086/323584}, \href
  {https://ui.adsabs.harvard.edu/abs/2001ApJS..137..279M} {137, 279}

\bibitem[\protect\citeauthoryear{{Moore}, {Katz}, {Lake}, {Dressler}  \&
  {Oemler}}{{Moore} et~al.}{1996}]{Moore_1996}
{Moore} B.,  {Katz} N.,  {Lake} G.,  {Dressler} A.,   {Oemler} A.,  1996,
  \mn@doi [\nat] {10.1038/379613a0}, \href
  {https://ui.adsabs.harvard.edu/abs/1996Natur.379..613M} {379, 613}

\bibitem[\protect\citeauthoryear{{Okabe}, {Futamase}, {Kajisawa}  \&
  {Kuroshima}}{{Okabe} et~al.}{2014}]{Okabe_2014}
{Okabe} N.,  {Futamase} T.,  {Kajisawa} M.,   {Kuroshima} R.,  2014, \mn@doi
  [\apj] {10.1088/0004-637X/784/2/90}, \href
  {https://ui.adsabs.harvard.edu/abs/2014ApJ...784...90O} {784, 90}

\bibitem[\protect\citeauthoryear{Owers et~al.,}{Owers
  et~al.}{2019}]{Owers_2019}
Owers M.~S.,  et~al., 2019, \mn@doi [The Astrophysical Journal]
  {10.3847/1538-4357/ab0201}, 873, 52

\bibitem[\protect\citeauthoryear{{Peng}, {Ho}, {Impey}  \& {Rix}}{{Peng}
  et~al.}{2010}]{Peng_2010}
{Peng} C.~Y.,  {Ho} L.~C.,  {Impey} C.~D.,   {Rix} H.-W.,  2010, \mn@doi [\aj]
  {10.1088/0004-6256/139/6/2097}, \href
  {https://ui.adsabs.harvard.edu/abs/2010AJ....139.2097P} {139, 2097}

\bibitem[\protect\citeauthoryear{{Plionis}}{{Plionis}}{1994}]{Plionis_1994}
{Plionis} M.,  1994, \mn@doi [\apjs] {10.1086/192104}, \href
  {https://ui.adsabs.harvard.edu/abs/1994ApJS...95..401P} {95, 401}

\bibitem[\protect\citeauthoryear{{Renzini}}{{Renzini}}{2006}]{Renzini_2006}
{Renzini} A.,  2006, \mn@doi [\araa] {10.1146/annurev.astro.44.051905.092450},
  \href {https://ui.adsabs.harvard.edu/abs/2006ARA&A..44..141R} {44, 141}

\bibitem[\protect\citeauthoryear{{Rom{\'a}n} \& {Trujillo}}{{Rom{\'a}n} \&
  {Trujillo}}{2017}]{Roman_2017}
{Rom{\'a}n} J.,  {Trujillo} I.,  2017, \mn@doi [\mnras] {10.1093/mnras/stx694},
  \href {https://ui.adsabs.harvard.edu/abs/2017MNRAS.468.4039R} {468, 4039}

\bibitem[\protect\citeauthoryear{{Ruiz-Lara} et~al.,}{{Ruiz-Lara}
  et~al.}{2018a}]{Ruiz_2018}
{Ruiz-Lara} T.,  et~al., 2018a, \mn@doi [\mnras] {10.1093/mnras/sty1112}, \href
  {https://ui.adsabs.harvard.edu/abs/2018MNRAS.478.2034R} {478, 2034}

\bibitem[\protect\citeauthoryear{{Ruiz-Lara} et~al.,}{{Ruiz-Lara}
  et~al.}{2018b}]{RL_2018}
{Ruiz-Lara} T.,  et~al., 2018b, \mn@doi [\mnras] {10.1093/mnras/sty1112}, \href
  {https://ui.adsabs.harvard.edu/abs/2018MNRAS.478.2034R} {478, 2034}

\bibitem[\protect\citeauthoryear{{Sales}, {Navarro}, {Pe{\~n}afiel}, {Peng},
  {Lim}  \& {Hernquist}}{{Sales} et~al.}{2020}]{Sales_2019}
{Sales} L.~V.,  {Navarro} J.~F.,  {Pe{\~n}afiel} L.,  {Peng} E.~W.,  {Lim} S.,
   {Hernquist} L.,  2020, \mn@doi [\mnras] {10.1093/mnras/staa854}, \href
  {https://ui.adsabs.harvard.edu/abs/2020MNRAS.494.1848S} {494, 1848}

\bibitem[\protect\citeauthoryear{{Schlafly} \& {Finkbeiner}}{{Schlafly} \&
  {Finkbeiner}}{2011}]{Schlafly_2011}
{Schlafly} E.~F.,  {Finkbeiner} D.~P.,  2011, \mn@doi [\apj]
  {10.1088/0004-637X/737/2/103}, \href
  {https://ui.adsabs.harvard.edu/abs/2011ApJ...737..103S} {737, 103}

\bibitem[\protect\citeauthoryear{Secker, Harris  \& Plummer}{Secker
  et~al.}{1997}]{Secker_1997}
Secker J.,  Harris W.~E.,   Plummer J.~D.,  1997, \mn@doi [Publications of the
  Astronomical Society of the Pacific] {10.1086/134018}, 109, 1377

\bibitem[\protect\citeauthoryear{{S\'ersic}}{{S\'ersic}}{1968}]{Sersic_1968}
{S\'ersic} J.~L.,  1968, {Atlas de galaxias australes}

\bibitem[\protect\citeauthoryear{{Smith}, {Lucey}, {Hudson}, {Allanson},
  {Bridges}, {Hornschemeier}, {Marzke}  \& {Miller}}{{Smith}
  et~al.}{2009}]{Smith_2009}
{Smith} R.~J.,  {Lucey} J.~R.,  {Hudson} M.~J.,  {Allanson} S.~P.,  {Bridges}
  T.~J.,  {Hornschemeier} A.~E.,  {Marzke} R.~O.,   {Miller} N.~A.,  2009,
  \mn@doi [\mnras] {10.1111/j.1365-2966.2008.14180.x}, \href
  {http://adsabs.harvard.edu/abs/2009MNRAS.392.1265S} {392, 1265}

\bibitem[\protect\citeauthoryear{Smith, Lucey, Price, Hudson  \&
  Phillipps}{Smith et~al.}{2011}]{Smith_2011}
Smith R.~J.,  Lucey J.~R.,  Price J.,  Hudson M.~J.,   Phillipps S.,  2011,
  \mn@doi [Monthly Notices of the Royal Astronomical Society]
  {10.1111/j.1365-2966.2011.19956.x}, 419, 3167–3180

\bibitem[\protect\citeauthoryear{{Taylor}}{{Taylor}}{2005}]{Taylor_2005}
{Taylor} M.~B.,  2005, in {Shopbell} P.,  {Britton} M.,   {Ebert} R.,  eds,
  Astronomical Society of the Pacific Conference Series Vol. 347, Astronomical
  Data Analysis Software and Systems XIV. p.~29

\bibitem[\protect\citeauthoryear{Terlevich, Caldwell  \& Bower}{Terlevich
  et~al.}{2001}]{Terlevich_2001}
Terlevich A.,  Caldwell N.,   Bower R.,  2001, \mn@doi [Monthly Notices of the
  Royal Astronomical Society] {10.1111/j.1365-2966.2001.04702.x}, 326, 1547

\bibitem[\protect\citeauthoryear{{Trujillo}, {Graham}  \& {Caon}}{{Trujillo}
  et~al.}{2001}]{Trujillo_2001}
{Trujillo} I.,  {Graham} A.~W.,   {Caon} N.,  2001, \mn@doi [\mnras]
  {10.1046/j.1365-8711.2001.04471.x}, \href
  {https://ui.adsabs.harvard.edu/abs/2001MNRAS.326..869T} {326, 869}

\bibitem[\protect\citeauthoryear{{Wittmann} et~al.,}{{Wittmann}
  et~al.}{2017}]{Wittmann_2017}
{Wittmann} C.,  et~al., 2017, \mn@doi [\mnras] {10.1093/mnras/stx1229}, \href
  {https://ui.adsabs.harvard.edu/abs/2017MNRAS.470.1512W} {470, 1512}

\bibitem[\protect\citeauthoryear{{Yagi}, {Koda}, {Komiyama}  \&
  {Yamanoi}}{{Yagi} et~al.}{2016}]{Yagi_2016}
{Yagi} M.,  {Koda} J.,  {Komiyama} Y.,   {Yamanoi} H.,  2016, \mn@doi [\apjs]
  {10.3847/0067-0049/225/1/11}, \href
  {http://adsabs.harvard.edu/abs/2016ApJS..225...11Y} {225, 11}

\bibitem[\protect\citeauthoryear{{Zabludoff} \& {Mulchaey}}{{Zabludoff} \&
  {Mulchaey}}{1998}]{Zabludoff_1998}
{Zabludoff} A.~I.,  {Mulchaey} J.~S.,  1998, \mn@doi [\apj] {10.1086/305355},
  \href {https://ui.adsabs.harvard.edu/abs/1998ApJ...496...39Z} {496, 39}

\bibitem[\protect\citeauthoryear{{Zaritsky} et~al.,}{{Zaritsky}
  et~al.}{2019}]{Zaritsky_2019}
{Zaritsky} D.,  et~al., 2019, \mn@doi [\apjs] {10.3847/1538-4365/aaefe9}, \href
  {https://ui.adsabs.harvard.edu/abs/2019ApJS..240....1Z} {240, 1}

\bibitem[\protect\citeauthoryear{{de Vaucouleurs}}{{de
  Vaucouleurs}}{1948}]{deV_1948}
{de Vaucouleurs} G.,  1948, Annales d'Astrophysique, \href
  {https://ui.adsabs.harvard.edu/abs/1948AnAp...11..247D} {11, 247}

\bibitem[\protect\citeauthoryear{{de Vaucouleurs}}{{de
  Vaucouleurs}}{1959}]{deV_1959}
{de Vaucouleurs} G.,  1959, Handbuch der Physik, \href
  {https://ui.adsabs.harvard.edu/abs/1959HDP....53..275D} {53, 275}

\bibitem[\protect\citeauthoryear{{van Dokkum} et~al.,}{{van Dokkum}
  et~al.}{2015}]{vanDokkum_2015}
{van Dokkum} P.~G.,  et~al., 2015, \mn@doi [\apjl]
  {10.1088/2041-8205/804/1/L26}, \href
  {http://adsabs.harvard.edu/abs/2015ApJ...804L..26V} {804, L26}

\bibitem[\protect\citeauthoryear{{van der Wel} et~al.,}{{van der Wel}
  et~al.}{2013}]{vanDerWell_2013}
{van der Wel} A.,  et~al., 2013, \mn@doi [\apjl] {10.1088/2041-8205/777/1/L17},
  \href {https://ui.adsabs.harvard.edu/abs/2013ApJ...777L..17V} {777, L17}

\makeatother
\end{thebibliography}

\appendix
\section{Comparison with the literature}
\label{sec:lit_cmp}
\begin{figure}
    \includegraphics[width=0.49\textwidth]{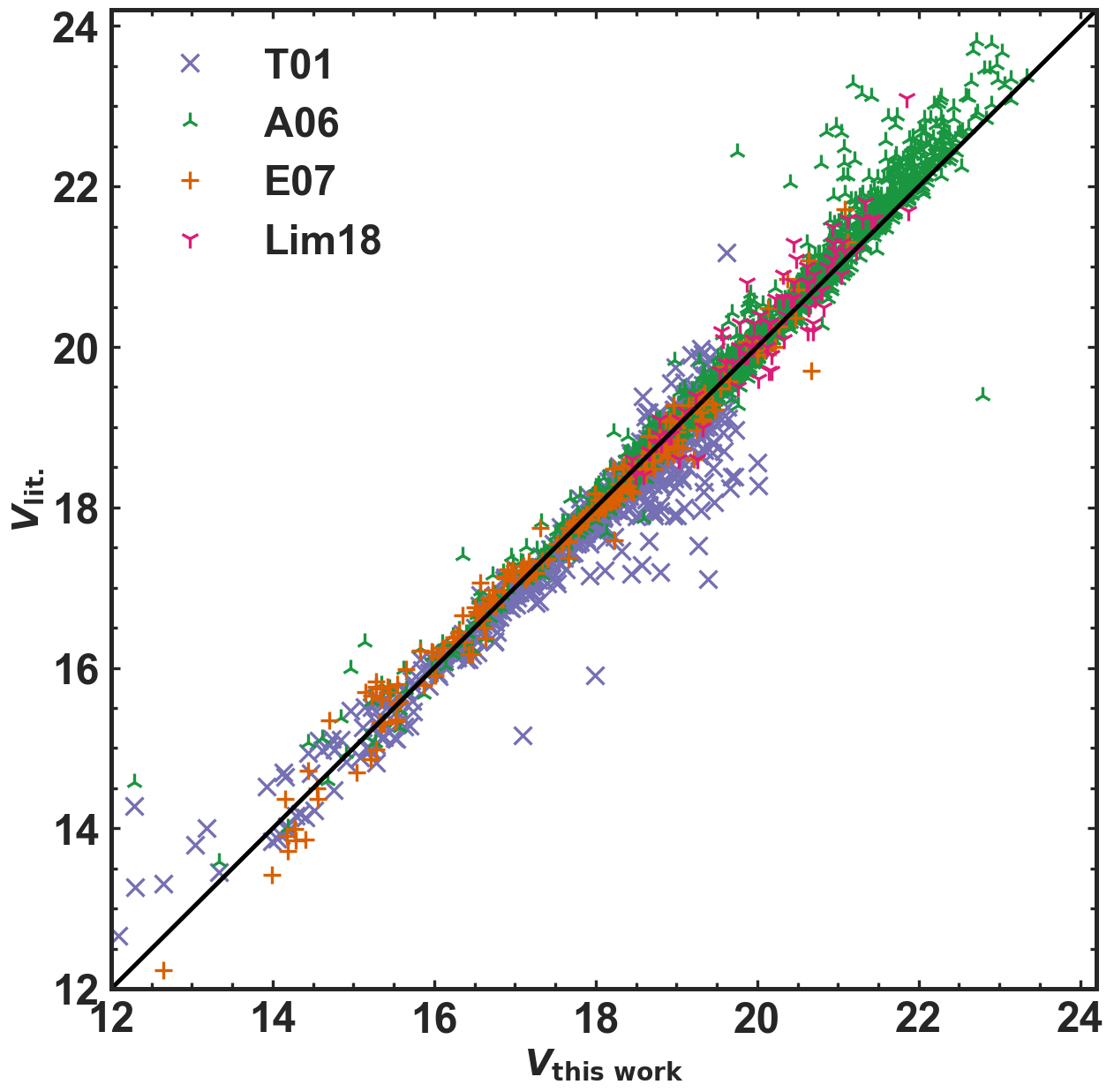}\hspace{0.01\textwidth}
    \includegraphics[width=0.49\textwidth]{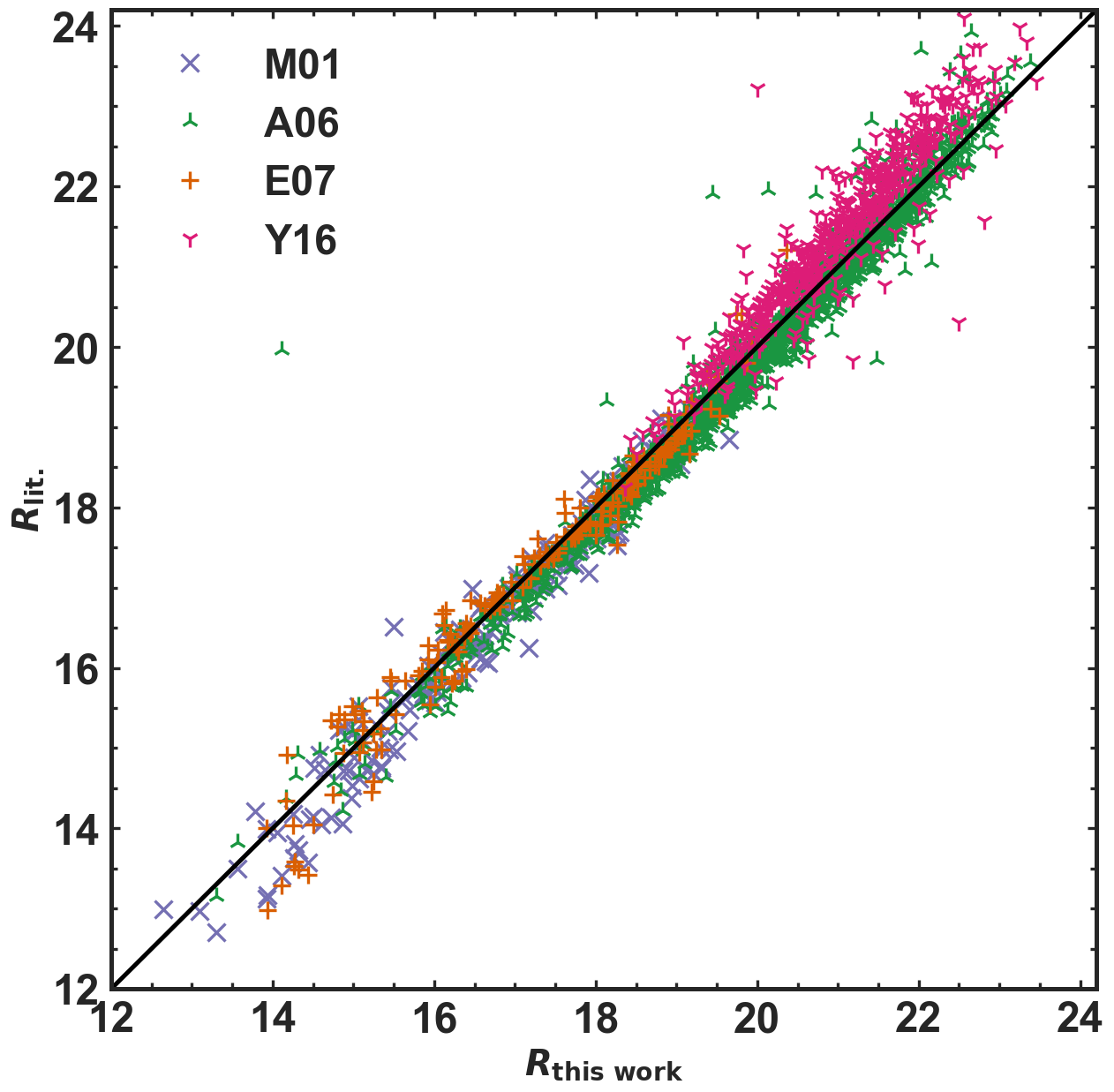}\hspace{0.01\textwidth}\\
	\caption{\label{fig:cmp_mag} Comparison of total magnitudes from our \texttt{GALFIT} analysis in $V$ and $R$-bands (\textit{left} and \textit{right panels}, respectively) with measurements from  \citet[][M01]{Mobasher_2001}, \citet[][T01]{Terlevich_2001}, \citet[][A06]{Adami_2006}, \citet[][E07]{Eisenhardt_2007}, \citet[][Y16]{Yagi_2016}, and \citet[][Lim18]{Lim_2018}. There is a very good agreement between our measurements and the literature over $\sim12$ magnitudes. Note that we report systematically brighter magnitudes for faint, low surface brightness galaxies due to the magnitude correction we applied as discussed in the text.}
\end{figure}

In this appendix we compare results from our \texttt{GALFIT} analysis with results from the literature, starting with total magnitudes. The Coma cluster has been well-studied in both $V$ and $R$ bands to various photometric depths and azimuthal coverage \citep[e.g.][etc.]{Terlevich_2001, Mobasher_2001, Adami_2006, Eisenhardt_2007, Yagi_2016}, mostly with \texttt{SExtractor} analysis and in the Vega magnitude system. We apply an average correction of $\sim0.3$ to account for the systematic offset between magnitude estimates from \texttt{SExtractor} analysis and \texttt{GALFIT} 2D--decomposition \citep[e.g.][etc]{Haussler_2007, vanDerWell_2013} and apply the appropriate corrections between the Vega and AB magnitude systems: $R_{\rm AB} = R_{\rm VEGA} + 0.21$ and $V_{\rm AB} = V_{\rm VEGA} + 0.02$ \citep{Blanton_2007}. We also include recent results from the F606W \textit{HST} study by \citet{Lim_2018} where a handful of Coma cluster galaxies were analysed with \texttt{GALFIT}. 
Figure~\ref{fig:cmp_mag} shows how our total magnitude measurements compare with several studies from the literature (all in AB magnitude system) over $\sim12$ magnitudes in both $V$ and $R$ bands.  

\begin{figure}
    \includegraphics[width=0.49\textwidth]{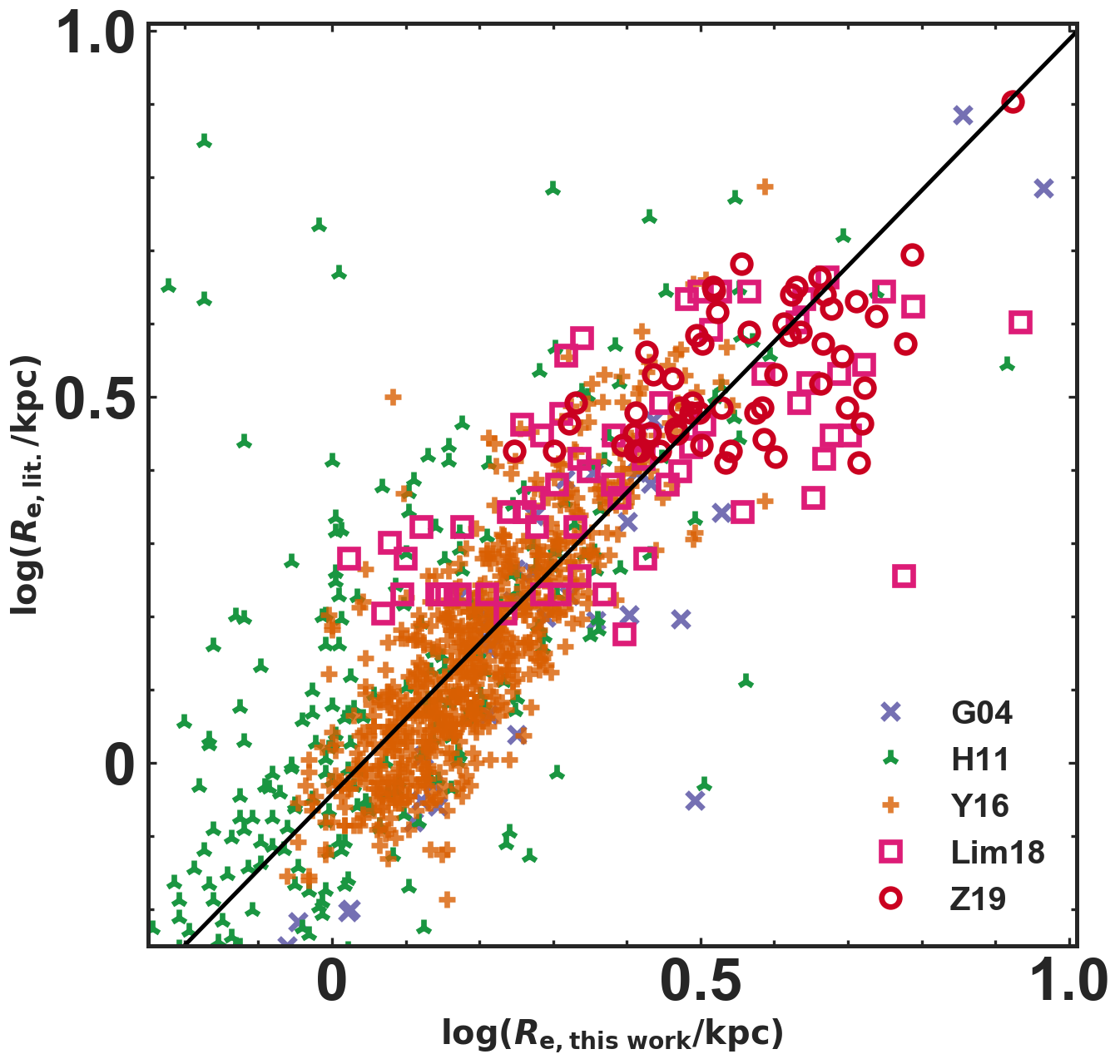}\hspace{0.01\textwidth}\\
	\caption{\label{fig:cmp_Re} Comparison of galaxy sizes, $\re$, with previous results from the literature as shown in the plot legend \citet[][G04]{Gutierrez_2004}, \citet[][H11]{Hoyos_2011}, \citet[][Y16]{Yagi_2016}, \citet[][Lim18]{Lim_2018}, and \citet[][Z19]{Zaritsky_2019}.}
\end{figure}

\begin{figure}
    \includegraphics[width=0.49\textwidth]{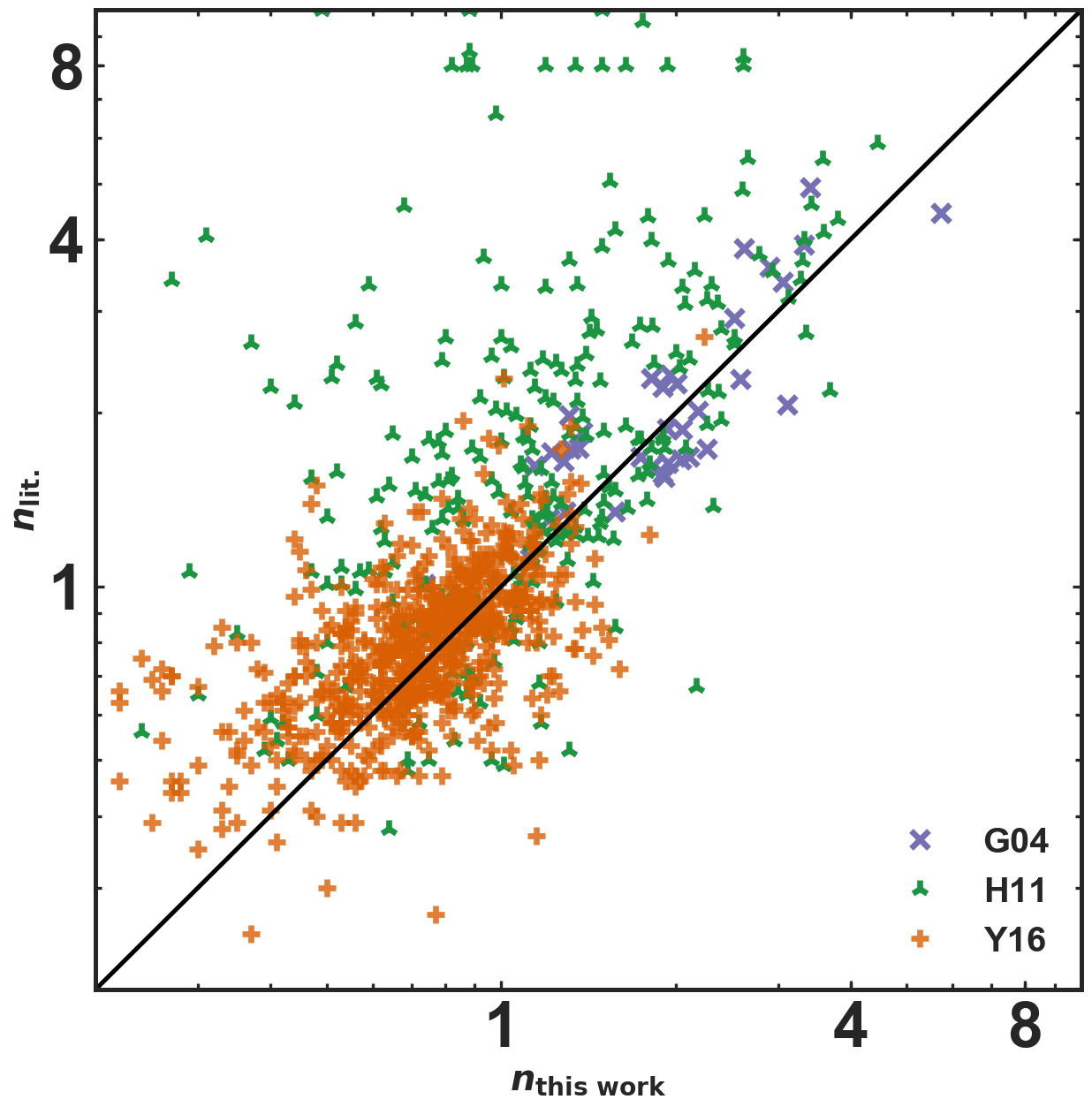}\hspace{0.01\textwidth}\\
	\caption{\label{fig:cmp_N} Same as in Figure~\ref{fig:cmp_Re} but showing galaxy \Sersic/ index, $n$.}
\end{figure}

\begin{figure}
    \includegraphics[width=0.49\textwidth]{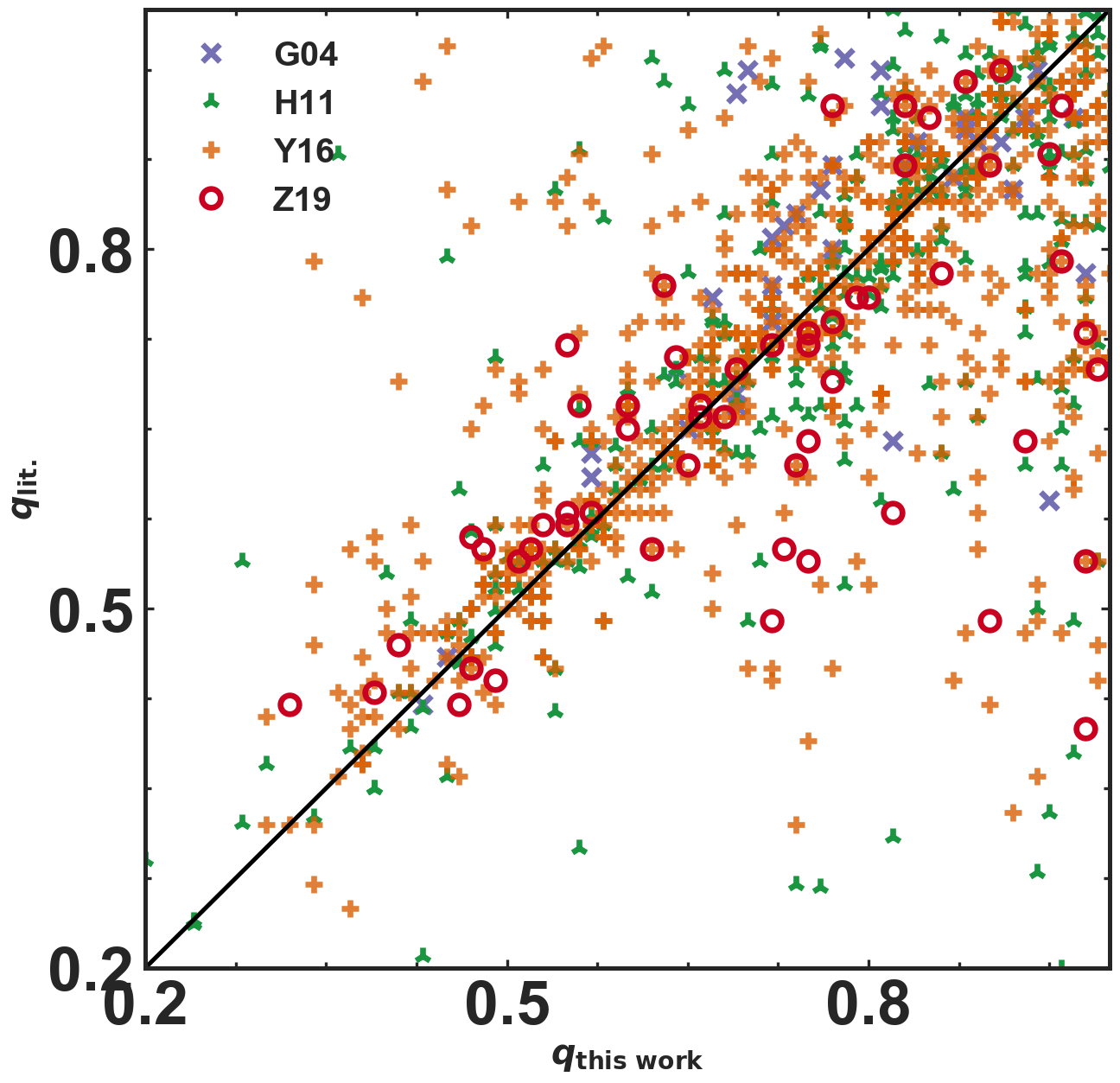}\hspace{0.01\textwidth}\\
	\caption{\label{fig:cmp_Q} Same as in Figure~\ref{fig:cmp_Re} but showing galaxy axial ratios, $q$.}
\end{figure}

\begin{figure}
    \includegraphics[width=0.49\textwidth]{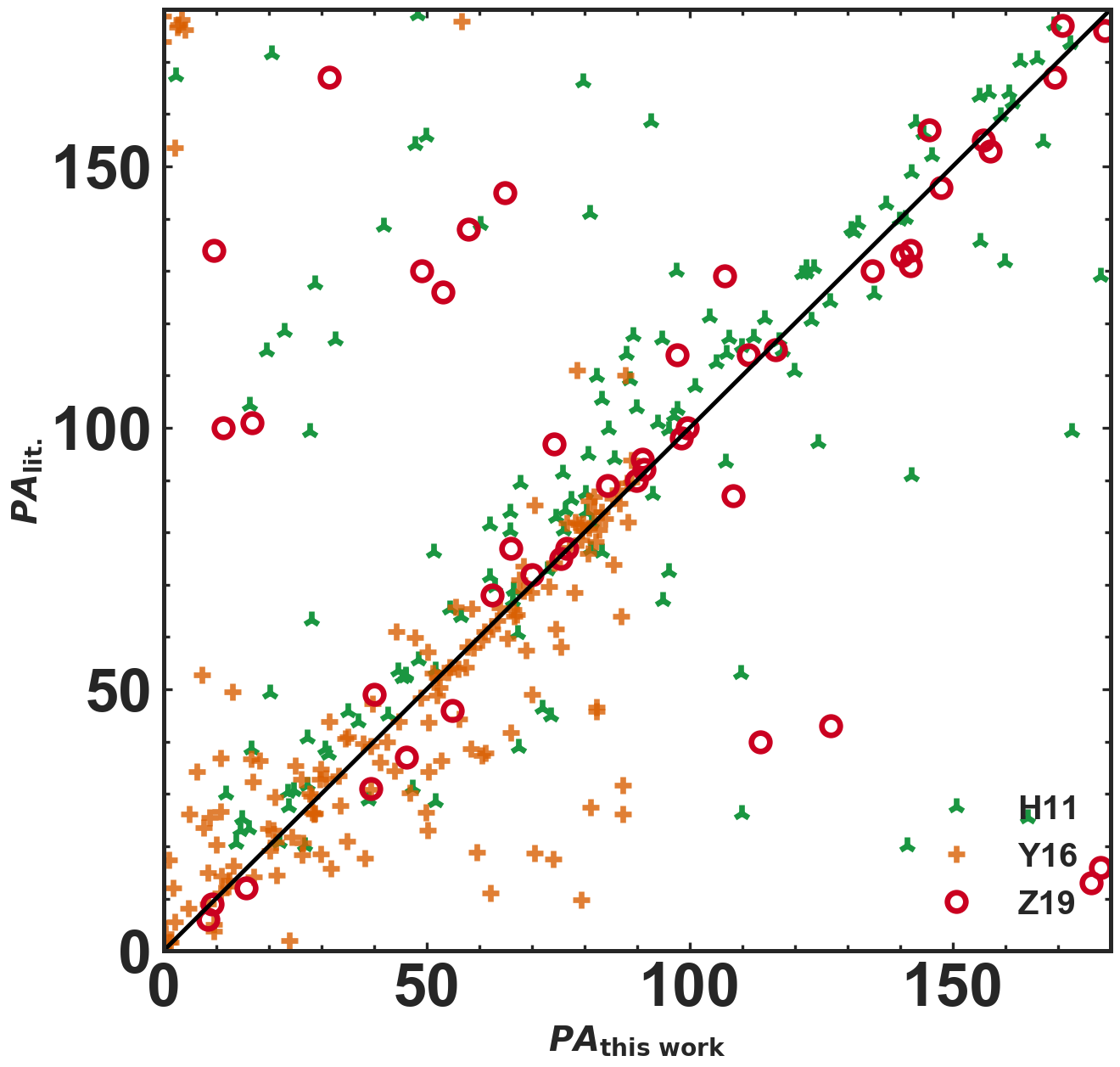}\hspace{0.01\textwidth}\\
	\caption{\label{fig:cmp_PA} Same as in Figure~\ref{fig:cmp_Re} but showing galaxy position angles, $PA$.}
\end{figure}

We also compare our measured effective radius $\re$ (Figure~\ref{fig:cmp_Re}), \Sersic/ index $n$ (Figure~\ref{fig:cmp_N}), axial ratio $q$ (Figure~\ref{fig:cmp_Q}), and position angle $PA$ (Figure~\ref{fig:cmp_PA}) with results from various literature sources including \citet{Gutierrez_2004, Hoyos_2011, Yagi_2016, Lim_2018, Zaritsky_2019}. Generally, our measured galaxy structural parameters show good agreement with previous works from the literature bearing in mind the various set of assumptions\footnote{For example, some authors fixed $n=1$ in their \texttt{GALFIT} analysis and have adopted various methods in dealing with the background sky, both factors which affect the values of the modelled parameters significantly.} and the different nature of the data we have compared with\footnote{Our comparison dataset include ground-based and \textit{HST} data in several photometric bands with varying resolutions.}

\bsp
\label{lastpage}
\end{document}